\documentclass[smallextended]{svjour3}       
\smartqed  

\usepackage{mathptmx}      
\usepackage{graphicx,psfrag}
\usepackage[usenames]{color}
\usepackage{amsmath}
\usepackage{amssymb}
\usepackage{natbib}

\newcommand{\etal}{\textit{et al.}}
\newcommand{\eq}[1]{\eqref{#1}}
\newcommand{\fig}[1]{figure~\ref{fig:#1}}

\newcommand{\Fig}[1]{Figure~\ref{fig:#1}}

\newcommand{\tab}[1]{Table~\ref{tab:#1}}

\newcommand{\const}{\mathrm{const}}
\renewcommand{\d}{\mathrm{d}}
\newcommand{\Df}[2]{\frac{\d{#1}}{\d{#2}}}
\newcommand{\df}[2]{\frac{\partial{#1}}{\partial{#2}}}
\newcommand{\E}[1]{\times10^{#1}}
\renewcommand{\O}[1]{\mathcal{O}\left(#1\right)}
\newcommand{\Real}{\mathbb{R}}

\newcommand{\uM}{\mu\mathrm{M}}
\newcommand{\s}{\mathrm{s}}

\newcommand{\NFKB}{NF-$\kappa$B}
\newcommand{\TNFA}{TNF$\alpha$}
\newcommand{\IKB}{I$\kappa$B}
\newcommand{\IKBA}{I$\kappa$B$\alpha$}

\newtheorem{thm}{Theorem}
\newtheorem{defn}{Definition}
\newcommand{\dom}[1]{\mathop{\mathrm{dom}}\left(#1\right)}

\newcommand{\NFkB}{p}
\newcommand{\IkBa}{q}
\newcommand{\nNFkB}{r}
\newcommand{\nIkBa}{s}
\newcommand{\tIkBa}{u}
\newcommand{\IKKn}{v}
\newcommand{\IKKa}{w}
\newcommand{\tA}{x}
\newcommand{\A}{y}
\newcommand{\pIkBaNFkB}{z}
\newcommand{\IKKi}{a}
\newcommand{\pIkBa}{b}
\newcommand{\nIkBaNFkB}{c}
\newcommand{\IkBaNFkB}{d}
\newcommand{\kv}{k_1}
\newcommand{\IK}{k_2}
\newcommand{\NF}{k_3}
\newcommand{\kaoa}{k_4}
\newcommand{\coa}{k_5}
\newcommand{\cta}{k_6}
\newcommand{\ctha}{k_7}
\newcommand{\cfa}{k_8}
\newcommand{\co}{k_9}
\newcommand{\ct}{k_{10}}
\newcommand{\cth}{k_{11}}
\newcommand{\cf}{k_{12}}
\newcommand{\kitha}{k_{13}}
\newcommand{\ketha}{k_{14}}
\newcommand{\kio}{k_{15}}
\newcommand{\keo}{k_{16}}
\newcommand{\kcoa}{k_{17}}
\newcommand{\kcta}{k_{18}}
\newcommand{\ktta}{k_{19}}
\newcommand{\kp}{k_{20}}
\newcommand{\kbA}{k_{21}}
\newcommand{\ka}{k_{22}}
\newcommand{\ki}{k_{23}}
\newcommand{\TR}{k_{24}}

\newcommand{\speed}{\lambda}    
\newcommand{\ssa}[2]{\overline #1^{#2}} 

\journalname{Journal of Mathematical Biology}

\begin{document}

\title{A method of `speed coefficients' for biochemical model reduction applied to the NF-kappaB system}

\subtitle{ 
\thanks{This study was supported 
in part by the Biotechnology and Biological Sciences Research Council 
(BBSRC) grants BBF0059381 and BBF5290031. Dr P. Paszek holds a BBSRC David Phillips 
Research Fellowship (BB/I017976/1).  }
}

\titlerunning{Minimal models of the NF-kappaB system}        

\author{Simon West \and Lloyd J. Bridge \and Michael R.H. White \and Pawel Paszek \and Vadim N. Biktashev  
}

\institute{
  S. West \at
  Institute of Integrative Biology, Crown Street, University of Liverpool, L69 7ZB\\
  Tel.: +151-795-4454\\
  \email{simon.west@liv.ac.uk} 
  \and
  L.J. Bridge \at
  Faculty of Life Science, University of Manchester, Oxford Road, Manchester, M13 9PT\\
  Tel.: +161-275-1834\\
  \email{lloyd.bridge@manchester.ac.uk}
  \and
  M. White \at
  Faculty of Life Science, University of Manchester, Oxford Road, Manchester, M13 9PT\\
  Tel.: +161-275-1834\\
  \email{mike.white@manchester.ac.uk}
  \and
  P. Paszek \at
  Faculty of Life Science, University of Manchester, Oxford Road, Manchester, M13 9PT\\
  Tel.: +161-275-1734\\
  \email{pawel.paszek@manchester.ac.uk}
  \and
  V. N. Biktashev \at
  College of Engineering, Mathematics and Physical Sciences,
  University of Exeter, North Park Road, Exeter EX4 4QF\\
  Tel.: +1392-72-6636\\
  \email{v.n.biktashev@exeter.ac.uk} 
}

\date{Received: date / Accepted: date}

\maketitle

\begin{abstract}
  The relationship between components of biochemical network and the
  resulting dynamics of the overall system is a key focus of
  computational biology. However, as these networks and resulting
  mathematical models are inherently complex and non-linear, the
  understanding of this relationship becomes challenging.  Among many
  approaches, model reduction methods provide an avenue to extract
  components responsible for the key dynamical features of the
  system. Unfortunately, these approaches often
    require intuition to apply. In this manuscript we propose a
  practical algorithm for the reduction of biochemical
  reaction systems using fast-slow asymptotics. This
  method allows the ranking of system variables according to how
  quickly they approach their momentary steady state, thus selecting
  the fastest for a steady state approximation. We applied this method
  to derive models of the Nuclear Factor kappa B network, a key
  regulator of the immune response that exhibits oscillatory dynamics.
  Analyses with respect to two specific solutions, which corresponded
  to different experimental conditions identified different components
  of the system that were responsible for the respective
  dynamics. This is an important demonstration of how reduction
  methods that provide approximations around a specific steady state,
  could be utilised in order to gain a better understanding of network
  topology in a broader context.  
\end{abstract}

\section{Introduction}
\label{intro}
\noindent Biological systems are inherently complex.  They are governed by a large number of functionally 
diverse components, which interact selectively and nonlinearly to achieve coherent outcomes \citep{Kitano}. 
Systems biology addresses this complexity by integrating biological experiments with computational 
methods, to understand how the components of a system interact and contribute to the biological function. 
However, the dynamical models that represent biological systems can often have high-dimensional state 
space and depend on a large number of parameters. Understanding the relationships between structure, 
parameters and function of such large systems is often a challenging and computationally intensive task. 

\begin{figure}[tbp]
\centerline{\includegraphics[width=1\textwidth]{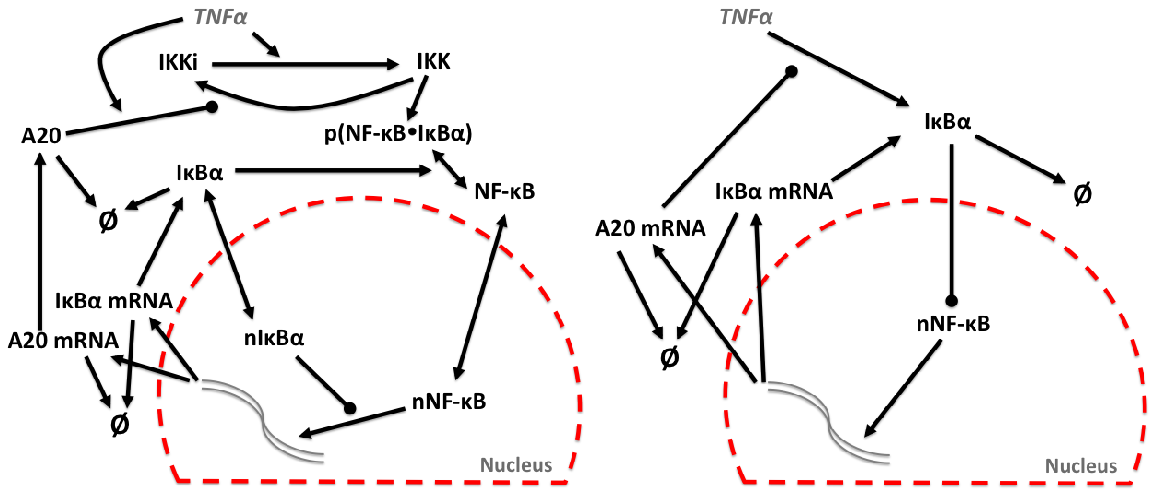}}
\caption{
Network diagram of the Simplified Model (derived from
\citealt{Ashall-etal-2009}) and the minimal model of the \NFKB\
system.  Time-dependent variables present in each model are depicted
with black colour. Pointed and round arrowheads represent activating
and inhibitory reactions, respectively. In unstimulated conditions
\NFKB\ is sequestered in the cytoplasm by association with \IKBA\
inhibitors. Stimulation with \TNFA\ (by changing $\TR=1$ from 0)
causes activation of the IKK kinase, and subsequently degradation of
\IKBA\ and translocation of free \NFKB\ to the nucleus. Nuclear \NFKB\
induces transcription of \IKBA\ and A20. Once synthesised \IKBA\ is
able to bind to \NFKB\ and return it to the cytoplasm, while A20
inhibits the IKK activity.
}
\label{fig:Fig1}
\end{figure}

One example of such a complex and high-dimensional system is the signalling network of the Nuclear 
Factor kappa B (\NFKB) transcription factor. \NFKB\ dynamics affects cell fate through the action of dimeric 
transcription factors that regulate immune responses, cell proliferation and apoptosis \citep{Hayden}. In 
unstimulated cells \NFKB\ is sequestered in the cytoplasm by association with the Inhibitor kappaB (\IKB) 
family of proteins. Upon stimulation with cytokines, such as Tumour Necrosis Factor $\alpha$ 
(\TNFA), the {\IKB}s are degraded releasing \NFKB\ to the nucleus 
where it activates the transcription of  over 300 target genes \citep{Hoffmann_circuitry}. Single cell fluorescence 
imaging has shown that upon continuous \TNFA\ stimulation \NFKB\ exhibits nuclear-to-cytoplasmic 
oscillations with a period of approximately 100 minutes \citep{Nelson}. This period is critical for maintaining 
downstream gene expression \citep{Ashall-etal-2009}. The oscillatory dynamics emerge through the interplay 
of a number of negative and positive feedback genes that are under the transcription control of 
NF-$\kappa$B. These, among others, include the \IKB\ and A20 inhibitors, and cytokines such as \TNFA\ (\fig{Fig1}) 
\citep{Hoffmann_circuitry}. In order to understand this intricate feedback regulation various mathematical 
models of the \NFKB\ signalling network have been proposed 
\citep{Hoffmann_2002,Lipniacki-etal-2004,Mengel-2012,Turner-2010}. However, the overall system is not fully resolved.

The large number of variables and biochemical reactions in dynamic
models, such as those of the \NFKB\ system, makes them analytically
intractable. Sensitivity analyses are often employed to understand
these models, assessing how individual parameters influence model
dynamics in a local and global context
\citep{Ihekwaba,Ihekwaba-2005,Rand-2008}. Model reduction approaches
provide a complementary avenue to extract the core reactions and
variables responsible for the key dynamical features of the
system. These include modularisation to break large systems down into
more tractable functional units \citep{Saez-Rodriguez-2004}.  However,
definition of a module becomes arbitrary, so this remains a heuristic
technique. Other techniques include using \emph{a posterori} analysis
and characteristic timescales. Based on error analysis, the former
method identifies, for different time intervals, the components of the
model required for accurate representation of the solution and uses
this information to guide model simplification
\citep{Whiteley-2010}. The latter utilises the fact that many
biological systems incorporate markedly different time-scales ranging
from seconds to hours.  Relevant approaches employ the use of
partial-equilibriums (PE), quasi-steady-state approximations (QSSA),
or grouping variables with equivalent time-scales
\citep{Krishna,Maeda-1998,Schneider-2000}, 
 see also \citep{Kutumova-etal-2013} and \cite{Radulescu-etal-2008}
for analysis of the NF-kappaB signalling.
These methods often rely on
intuition to identify the small parameters that
allow the successive reduction steps, and a standard problem for
perturbation methods is that in reality the small parameters are
never infinitely small and one needs somehow to assess whether they
are small enough for any particular purpose, that is, additional
  accuracy control is required.
Algorithmic
approaches to identification of small parameters been proposed.  For
instance, Computational Singular Perturbation (CSP) method is an
iterative procedure, based on identification of the fast modes
through the analysis of of the eigenvalues of the Jacobian matrix
\citep{Lam-Goussis-1994}, %
see also~\citep{Kourdis-etal-2013}
for the asymptotic analysis of the NF-kB dynamics.
Other methods exploiting the eigenvalues of the
Jacobian are the Intrinsic Low-Dimensional Manifolds (ILDM) method
by \cite{Maas-Pope-1992} and a more refined Method of
Invariant Manifold by \cite{Gorban-Karlin-2003}. Comparison and analysis of these
methods can be found in \citep{Zagaris-etal-2004}. 
Although these methods are more advanced
that the classical PE and QSSA techniques, they are also more
technically challenging than their predecessors. QSSA
methods retain original variables and parameters. Alternative methods,
such as the Elimination of Nonessential Variables (ENVA) method
described in~\citep{Dano-etal-2006} exploit searches through
lower-dimensional models of reduced networks for a minimal
mathematical model which will reproduce a desired dynamic behaviour of
the full model. Such a systematic reduction method has the advantage
of requiring neither knowledge of the minimal structures, nor
re-parameterisation of the retained lumped model components. Indeed,
application of model reduction methods which are algorithmic rather
than necessarily biologically intuitive can clearly reveal model
sub-structures which control basic system dynamics.

In this manuscript we use a simple algorithmic QSSA
approach for the reduction of biochemical reaction systems
using a heuristic that is likely to be widely applicable
to this sort of systems.  We define
``speed coefficients'' that enable ranking variables according to how
quickly their approach their momentary steady-state. This allows a
straightforward choice of variables for elimination by QSSA
application at each step of the algorithm, while preserving dynamic
characteristics of the system. We use this method to derive reduced
models of the \NFKB\ signalling network. Our analysis identifies the
key feedback components of the system responsible for \NFKB\
dynamics. Further, reduction of the \NFKB\ model around different
solutions (corresponding to different experimental protocols) revealed
specific components of the IKK signalling module responsible for
generation of the respective dynamics. This demonstrates the
application of an essentially local technique can be used to infer
information about the system in a larger context, ultimately providing
a better understanding of the \NFKB\ signalling network.

\section{Methods}
\label{sec:Methods}

\subsection{Perturbation theory for fast-slow systems}
\label{sec:perturbation}

The application of steady-state approximation to biochemical reaction
systems typically argues that some of the reagents are highly
reactive, so are used as quickly as they are made. Therefore, after
the initial transient phase, the concentration of such a reagent is
always close to what would be its steady-state as long as
concentrations of other reagents were maintained constant. In the
simplest form, this means that in the kinetic equations, the
corresponding rate of change can be set to zero.  This provides a
general procedure for simplifying biochemical systems, based on the
difference of characteristic time-scales. Practical
application of this idea dates back at least to
\citet{Briggs-Haldane-1925}. More recent reviews and textbook
expositions can be found e.g. in
\citep{Klonowski-1983,Segel-Slemrod-1989,Volpert-Hudjaev-1985,Yablonskii-etal-1991}.
The
basic mathematical justification of the formal procedures stems from
the seminal work by Tikhonov~\citeyearpar{Tikhonov-1952}.  It is
formulated for systems which involve small parameter $\epsilon$ in the
form
\begin{equation}                                  \label{slow-fast}
  \begin{split}
    \Df{x}{t} &= f\left(x,z,t\right),            
\\
    \epsilon\Df{z}{t} &= g\left(x,z,t\right),    
  \end{split}
\end{equation}
where $x$ is a vector of slow variables and $z$ is the vector of fast
variables. In the limit $\epsilon\to0$, the system~\eq{slow-fast} becomes
\begin{equation}                                  \label{slow-limit}
  \begin{split}
  \Df{x}{t} &= f\left(x,z,t\right),               
\\
          z &= \phi\left(x,t\right),              
 \end{split}
\end{equation}
where $\phi\left(x,t\right)$ is the solution of
$g\left(x,z,t\right)=0$.  If $\epsilon$ is small, the solutions to the
original system \eq{slow-fast} may be expected to differ from solutions of
\eq{slow-limit} only slightly. For an
initial-value problem for a finite time interval this is guaranteed
by the following:
\begin{thm}\label{Tikhonov}
  Let the right-hand sides of systems~\eq{slow-fast} and
  \eq{slow-limit} be sufficiently smooth so solutions to initial
  value problems exist and are unique.  Let $x=X(t;\epsilon)$,
  $z=Z(t;\epsilon)$, $t\in[0,T]$, $T>0$ be a solution of the
  system~\eq{slow-fast} with initial condition $X(0;\epsilon)=x_0$,
  $Z(0;\epsilon)=z_0$, and $x=\bar{X}(t)$ be a solution to the
  system~\eq{slow-limit} with initial condition
  $\bar{X}(0)=x_0$. Consider also the ``attached'' system,
  \begin{equation}
    \Df{z}{s} = g\left(x,z,t\right),          \label{fast}
  \end{equation}
  depending on $x$ and $t$ as parameters. Let $z=\phi(x,t)$ be a function
  defined on an open set containing the trajectory $\{(\bar{X}(t),t),
  t\in[0,T]\}$, such that $z=\phi(\bar{X}(t),t)$ is an isolated,
  Lyapunov stable and asymptotically stable equilibrium of \eq{fast} for the
  corresponding $x=\bar{X}(t)$ and any $t\in[0,T]$. Finally, assume
  that $z_0$ is within the basin of attraction of the equilibrium
  $\phi(x_0,0)$ of system~\eq{fast} at $x=x_0$, $t=0$. 
  Then for any $t\in(0,T]$,
  \[
  \lim\limits_{\epsilon\to0} X(t;\epsilon) = \bar{X}(t), 
  \qquad
  \lim\limits_{\epsilon\to0} Z(t;\epsilon) = \phi(\bar{X}(t)).
  \]
\end{thm}

This theorem is a special case of Theorem~1
of~\cite{Tikhonov-1952}. In fact, the solution of the full
system~\eq{slow-fast} can be considered as consisting of two parts:
the initial transient, approximately described by~\eq{fast}, with
$s=\epsilon t$, and $x\approx x_0$, which is followed by the
long-term part, approximately described by the solution
$x=\bar{X}(t)$, $z=\phi(x)$. However the duration of the transient is
$\O{\epsilon}$ so for any fixed $t>0$ and sufficiently small
$\epsilon$, the initial transient will have expired by the time $t$,
hence the limit.

A limitation of the above result is that it gives only pointwise
convergence in $\epsilon$ so it does not answer the questions about
the behaviour of trajectories as $t\to0$ at a fixed $\epsilon$. There
were later extensions of this work, relieving this limitation. In
this paper we will be looking at periodic solutions, so the following
result is relevant to us:

\begin{thm}\label{Anosov}
  In addition to the assumptions of Theorem~\ref{Tikhonov}, suppose that
  the slow system~\eq{slow-limit} has a periodic solution with
  period $P_0$, that is $x=\tilde{X}(t)$: $\tilde{X}(t+P_0)\equiv \tilde{X}(t)$, and this solution is
  stable in the linear approximation. Then the full
  systems~\eq{slow-fast} have an ($\epsilon$-dependent) family of
  periodic solutions with periods $P(\epsilon)$ such that
  $\lim_{\epsilon\to0}P(\epsilon)=P_0$ and the corresponding orbits
  lie in a small vicinity of $(\tilde{X}(t),\phi(\tilde{X}(t)))$ for small
  $\epsilon$. Moreover, the periodic orbits and the period depend of
  $\epsilon$ smoothly.
\end{thm}
This theorem is a special case of Theorem~5 of~\citep{Anosov-1960}. 

When the approximation of the solution of the full system by that of
the slow system is insufficient in itself, it can be improved by
considering higher-order corrections in $\epsilon$. The mathematical
justification of that procedure is based on the results about
smoothness of the dependence of solutions of the full system on
$\epsilon$, see e.g.~\cite{Vasilieva-1952}. A very influential
continuation of these works with important generalizations, under
a currently popular name of ``geometric perturbation theory'', has been
done by \citet{Fenichel-1979}. Below we present a simple illustration
of the method, directly applicable to our situation.

\subsection{Identification of small parameters: parametric embedding}
\label{sec:embedding}

In the real-life kinetic equations it is not always obvious which
reagents can be suitable for the QSSA.  To identify such reagents, we
follow the formal method of ``parametric embedding''
\citep{Suckley-Biktashev-2003,Biktasheva-2006}.

\begin{defn}
  We will call a system %
  \[
  \dot{u} = F(u;\epsilon), \qquad u\in\Real^{d},
  \]
  depending on parameter $\epsilon$, 
  a \emph{1-parametric embedding} of a system
  \[
  \dot{u} = f(u), \qquad u\in\Real^{d},
  \]
  if $f(u)\equiv F(u,1)$ for all $u\in\dom{f}$. 
  If the limit $\epsilon\to0$ is concerned then we call it a
  \emph{asymptotic embedding}. If 
  a 1-parametric embedding has a form \eq{slow-fast}, we call it a
  \emph{Tikhonov embedding}.
\end{defn}

The typical use of this procedure has the form of a replacement of a
small constant with a small parameter. If a system contains a
dimensionless constant $a$ which is ``much smaller than 1'', then
replacement of $a$ with $\epsilon a$ constitutes a 1-parametric
embedding; and then the limit $\epsilon\rightarrow0$ can be
considered. In practice, constant $a$ would more often be replaced
with parameter $\epsilon$, but mathematically, in the context of
$\epsilon\rightarrow0$ and $a=\const\ne0$ this, of course, does not
make any difference from $\epsilon a$. This explains the
paradoxical use of a zero limit for a parameter whose true
value is one.

In some applications, the ``small parameters'' appear naturally and
are readily identified. However, this is not always the case, and
in complex nonlinear systems
asymptotic analysis may require this procedure of parametric
embedding, i.e. introduction of small parameters artificially. It is
important to understand, that there are infinitely many ways a given
system can be parametrically embedded, as there are infinitely many
ways to draw a curve $F(u;\epsilon)$ in the functional space given
the only constraint that it passes through a given point,
$F(u;1)=f(u)$.  In terms of asymptotics, which of the embeddings is
``better'' depends on the qualitative features of the original
systems that need to be represented, or classes of solutions that
need to be approximated. Some examples of different Tikhonov
embeddings of a simple cardiac excitation model can be found
in~\cite{Suckley-Biktashev-2003}, and non-Tikhonov embedding of the
same in~\cite{Biktashev-Suckley-2004}, and some of those examples
are better than others in describing particularly interesting
features of cardiac action potentials. 

If a numerical solution of the system can be found easily, then there
is a simple practical recipe: to look at the solutions of the
embedding at different, progressively decreasing values of the
artificial small parameter $\epsilon$, and see when the features of
interest will start to converge. If the convergent behaviour is
satisfactorily similar to the original system with $\epsilon=1$, the
embedding is adequate for these features.  

To summarize, we claim that identification of small parameters in a
given mathematical model with experimentally measured functions and
constants will, from the formal mathematical viewpoint,
always be arbitrary (even though in the simplest cases there may be
such a natural choice that this ambiguity is not even realized by
the modeller), and ``validity'' of such identification can be defined
only empirically: if the asymptotics describe the required class of
solutions sufficiently well. The rare exceptions are when the
asymptotic series are in fact convergent and the residual terms can be
estimated a priori. A cruder (and less reliable) estimate of
the error of an asymptotic can be obtained through the analysis of the
higher-order asymptotics, see e.g.~\cite{Turanyi-etal-1993}; more
about it later.

In this paper, we restrict consideration to Tikhonov
embeddings~\eq{slow-fast}. The simplest version of the above recipe
results in the straightforward procedure: compare the solution
of the full system with the solution where the putative fast
variable has been replaced by its quasistationary value. In terms of
the ``numerical embedding'', this means a short-cut: considering
values $\epsilon=1$ and $\epsilon=0$ instead of a (or as a very
short) sequence of values of $\epsilon$ converging to 0. Although
sometimes we have indeed studied several values of $\epsilon$, we
shall always present only $\epsilon=1$ and $\epsilon=0$ results, to
avoid cluttering the graphs.

\subsection{Speed coefficients}
\label{sec:speed}

It follows from the above discussion that the
``numerical embedding'' procedure could be applied to any of the
dynamic variables, and those whose adiabatic elimination would cause
the smallest changes in the solution, could be taken as the fastest. In
practice, for a large system, this exhaustive trial and error
procedure may be too laborious. We employ a
simple heuristic method to identify the candidates for the fastest
variables.

We describe it in terms of a
generic system of $N$ ordinary differential equations (ODEs),
\begin{equation}
  \Df{x_i}{t} = f_{i}\left(x_{1},\dots ,x_{N}\right), \qquad i=1,\dots N.
                                                  \label{generic}
\end{equation}

 We define the ``speed coefficients'' for each dynamic variable $x_i$
as
\begin{equation}
  \speed_i \left(x_{1},...,x_{N}\right) = \left|\df{f_i}{x_i}\right|.
                                                  \label{speeds}
\end{equation}
\begin{figure}[tbp]
\centerline{\includegraphics[width=0.6\textwidth]{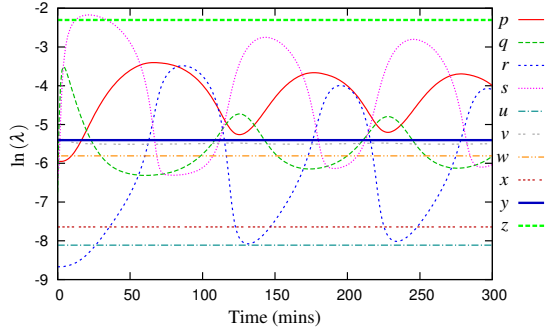}}
\caption{
Semi-logarithmic plot of speed coefficients for the Simplified Model
(SM).  A larger speed coefficient means the variable is approaching
its steady state faster. These coefficients identify variable
$\pIkBaNFkB$ as the fastest, and therefore the most appropriate
candidate for elimination.
}
\label{fig:Fig2}
\end{figure}
By definition, these coefficients depend on the dynamic variables, 
or, for a selected solution, they depend on time $t$. 
They can be used to rank the variables according to how quickly they
approach their momentary steady-states (\fig{Fig2}).

It is very essential to understand that with the
exception of relatively trivial cases, the most adequate choice of
embedding will depend on the type of solutions that are of interest
for the particular application at hand, because in a nonlinear
system, what is ``small'' and what is ``large'' may be significantly
different in different parts of the phase space. A simple but very
instructive example illustrating this point is considered
by~\citet[Section A]{Lam-Goussis-1994}, where the meaning of
fast/slow changes depending on initial conditions and on what part
of the solution is considered. Our practical approach is that we
start from one particular solution, which is selected in such a way
that to be sufficiently representative for the class of solutions
that are of interesting to a particular application. An obvious
extension would be selection of a representative set of solutions;
however for the illustration of the method, one is enough.  As
follows from the above, the task of selecting such solutions is
inevitably the responsibility of the investigator who is going to
apply the method and use the resulting reduced system. In the
particular models we consider here this task is relatively
straightforward, as the long-term behaviour is more or less the same
for any physiologically sensible initial conditions. For elimination
of any further  
ambiguity we have adopted a rule that we would select for elimination
the variable that is fastest at its slowest. That is, for each
variable we find the minimal value of its speed coefficient over the
simulated time interval, and then select the variable which has the
highest value of the minimal speed among other variables.

Note that our heuristic procedure only uses partial information about
the system (only the diagonal elements of the Jacobian, and only its
minimal value along only one/a few solution(s)),
but it is only used for preliminary selection of variables for
reduction. Therefore, the actual success of reduction is established
by comparison of the reduced and the original system, within the
``numerical embedding'' procedure described above. In the test cases
presented in this paper, this proof has always been successful, if
sometimes with first-order corrections. However one cannot exclude the
possibility that high relative values of the non-diagonal elements of
the Jacobian and/or its strong variations over the representative
solutions may force the change of the candidate for reduction, or QSSA
may be inapplicable in principle. As an extreme
example, consider a subsystem: $\dot{x}=Ay$, $\dot{y}=-Ax$, which
has zero diagonal Jacobian elements, so would be classed as
(infinitely) slow, yet for large $A$ its treatment as such within a
larger system would produce wrong results, as in fact $x$ and $y$
will fastly oscillate. For the (bio-)chemical kinetics this sort of
behaviour is, however, not very likely, at least at the level of
elemetary reactions; see e.g. the discussion
in~\citep[p. 165]{Turanyi-etal-1993}.  On the other hand, this
fastly oscillating subsystem is not appropriate for Tikhonov style
treatment anyway, and requires averaging in Krylov-Bogolyubov style
instead; whereas if a system \emph{does} have the form
\eq{slow-fast} and satisfies the assumptions of
Theorem~\ref{Tikhonov}, then the eigenvalues of the Jacobian block
$\epsilon^{-1}\partial{g}/\partial{z}$ have negative real parts and
are of the order of $\epsilon^{-1}$, so its diagonal elements are likely to be large
(and negative) --- although, of course, counter-examples can be
invented.

Finally, we note again that the choice of variables for reduction
may depend on the class of solutions of interest, which in our
approach will be done via the choice of representative solution. In
sections~\ref{sec:Ashall-cont} and \ref{sec:Ashall-pulse} we consider
two different classes of solution in the same full model, which give
two different reduced models.

\subsection{The  model reduction algorithm}
\label{sec:algorithm}

Based on Tikhonov's and Anosov's theorems and the definition of the
speed coefficients we can define a general method for reducing the
dimension of a biochemical reaction system. We illustrate the method using an example where
the right-hand side of an ordinary differential
equation for a fast variable is linear with respect to the same variable. Suppose the variable $x_j$
has been identified as the fast variable in the system
\eq{generic}.  With account of
the artificial small parameter, this gives
\begin{equation}
  \epsilon\Df{x_j}{t}=\alpha_j(t)-\beta_j(t)x_j ,   \label{quasilin} 
\end{equation}
where coefficients $\alpha_j(t)$ and $\beta_j(t)$ are presumed to
depend on time via other dynamic variables. 
We look for a solution in the form of an asymptotic series 
$x_j=x_j^0+\epsilon x_j^ 1+\epsilon^2x_j^ 2+\O{\epsilon^3}$. Substituting this into
\eq{quasilin} gives 
\begin{equation}
  \epsilon\dot{x}_j^0+\epsilon^2\dot{x}_j^1+\epsilon^3\dot{x}_j^2
  = \alpha_j-\beta_j x_j^0-\epsilon\beta_j x_j^1-\epsilon^2\beta_j x_j^2
  + \O{\epsilon^{3}} .                             \label{asseries}
\end{equation}
The simplest approximation for $x_j$ is obtained by considering the
terms in \eq{asseries} proportional to $\epsilon^0$, 
\begin{equation}
  0 = \alpha_j(t)-\beta_j(t)x^0_j,                     \label{order0}
\end{equation}
which results in the zeroth-order QSSA for variable $x_j$: 
\begin{equation}
  \ssa{x_j}{0} = x_j^{0}=\frac{\alpha_j(t)}{\beta_j(t)} .        \label{xj0}
\end{equation}
This approximation $x_j^0$ is then substituted into the original
system of equations for the variable $x_j$. If the variable is
sufficiently fast then this steady-state expression should be a good
approximation of the fast variable and the substitution will cause
minimal change to the solution.

In general, the zeroth-order QSSA provides a reasonable approximation of
the original variable. However, if such approximation is not good
enough, it can be improved by calculating an additional
correction term. To do this we consider terms in \eq{asseries}
proportional to $\epsilon^1$:
\begin{equation}
  \epsilon\dot{x}_j^0 = -\epsilon\beta_j x_j^1 .      \label{order1}
\end{equation}
Substituting our earlier result \eq{xj0} into equation \eq{order1} and
solving for $x_j^{1}$ gives the first-order correction in the form
\begin{equation}
  x_j^1 = -\frac{1}{\beta_j}\dot{x}_j^0 
  = \frac{\alpha_j\dot{\beta_j}-\beta_j\dot{\alpha_j}}{\beta_j^{3}}.
                                                  \label{xj1}
\end{equation}
This results in the first-order QSSA $\ssa{x_j}{1}=x_j^0+\epsilon x_j^1$ in the form
\begin{equation}
  \ssa{x_j}{1} = \frac{\alpha_j(t)}{\beta_j(t)} + \frac{
    \alpha_j(t)\dot{\beta_j}(t)-\beta_j(t)\dot{\alpha_j}(t)
  }{
    \beta_j^{3}(t)
  }                      ,                         \label{ssaxj1}
\end{equation}
since the original problem corresponds to $\epsilon=1$. Note
that the value of the first-order correction, or its estimate,
can be used as an estimate of the accuracy of the leading-term
approximation; roughly speaking, this is the idea behind the accuracy
estimate used in~\cite{Turanyi-etal-1993}.

So our method can then be formulated into a general algorithm to reduce
the dimension of a biochemical system defined by ordinary differential
equations. The algorithm reads:
\begin{enumerate}
\item Using numerical methods, find a representative solution of the
  system of ODEs for the chosen time interval.
\item Calculate the expressions for the speed coefficients ($\speed$'s),
  using equation \eq{speeds} from the system of ODEs
  (this can be assisted by a symbolic calculations software, e.g. Maple).
\item Substitute the numerical solution of the system into the expressions for
  the $\speed$'s to find the speed for each variable at each time
  point.
\item Plot the speed coefficients vs. time and identify the fastest
  variable (at its slowest).
\item Calculate the expression for the zeroth-order QSSA using
  \eq{xj0}. 
\item Substitute this QSSA into the system of ODEs to eliminate the
  fastest variable, thus obtaining a reduced system.
\item Compare the solution of the reduced system with the solution of
  the original system. 
\item \emph{If the zeroth-order QSSA is insufficient to maintain a
    suitable accuracy, calculate the first-order QSSA using equation
    \eq{ssaxj1}}.
\item Repeat the above process for the new reduced system.
\end{enumerate}

\section{Minimal model of the \NFKB\ system in response to continuous \TNFA\ input}
\label{sec:Ashall-cont}

\begin{table}[tbp]
 \caption{%
    Variables and parameters: their names as in \citet{Ashall-etal-2009},
    short names adopted here, and values.  The initial conditions were
    obtained by equilibrating the system without stimulus ($\TR=0$),
    with $\IKKn$ and $\nNFkB$ set to $\IK$ and $\NF\kv$, respectively, and other
    variables set to 0.
 }
 \label{tab:1}
\resizebox{\textwidth}{!}{\begin{minipage}{1.15\textwidth}
\newcommand{\fmt}{|l|l|l|}
\newcommand{\blank}{&&}
\begin{tabular}{\fmt}
\hline\multicolumn{3}{|c|}{\textbf{Variables} ($\uM$)} \\\hline
$NF\kappa B$       & $\NFkB$  & $3.81\E{-3}$ \\\hline 
$I\kappa B\alpha$  & $\IkBa$  & $1.58\E{-2}$ \\\hline 
$nNF\kappa B$      & $\nNFkB$ & $9.79\E{-3}$ \\\hline 
$nI\kappa B\alpha$ & $\nIkBa$ & $5.44\E{-3}$ \\\hline 
$tI\kappa B\alpha$ & $\tIkBa$ & $2.07\E{-5}$ \\\hline 
$IKKn$             & $\IKKn$  & $0.08$       \\\hline 
$IKKa$             & $\IKKa$  & $0$          \\\hline 
$tA20$             & $\tA$    & $6.46\E{-6}$ \\\hline 
$A20$              & $\A$     & $7.19\E{-4}$ \\\hline 
$pI\kappa B\alpha\circ NF\kappa B$  
               & $\pIkBaNFkB$ & $0$          \\\hline 
$IKKi$             & $\IKKi$  & $0$          \\\hline 
$pI\kappa B\alpha$ & $\pIkBa$ & $0$          \\\hline 
$nI\kappa B\alpha\circ nNF\kappa B$
               & $\nIkBaNFkB$ & $7.95\E{-4}$ \\\hline 
$I\kappa B\alpha\circ NF\kappa B$
                & $\IkBaNFkB$ & $7.30\E{-2}$ \\\hline 
\end{tabular}
\begin{tabular}{\fmt}
\hline\multicolumn{3}{|c|}{\textbf{Parameters}} \\\hline
$kv$    & $\kv$    & $3.3$          \\\hline
IKK$^*$\footnote[0]{$^*$ IKK$=\IKKn+\IKKa+\IKKi$ (conserved quantity)}
        & $\IK$    & $0.08\,\uM$         \\\hline
\NFKB$^\dagger$\footnote[0]{%
 $^\dagger$\NFKB $=\NFkB+\IkBaNFkB+\pIkBaNFkB+\frac1\kv(\nNFkB+\nIkBaNFkB)$
 (conserved quantity)
}
        & $\NF$    & $0.08\,\uM$         \\\hline
$ka1a$  & $\kaoa$  & $0.5\,(\uM\s)^{-1}$ \\\hline
$c1a$   & $\coa$   &  $1.4\E{-7}\,(\uM\s)^{-1}$          \\\hline
$c2a$   & $\cta$   & $0.5\,\s^{-1}$         \\\hline
$c3a$   & $\ctha$  & $0.0003\,\s^{-1}$      \\\hline
$c4a$   & $\cfa$   & $0.0005\,\s^{-1}$      \\\hline
$c1$    & $\co$    & $1.4\E{-7}(\uM\s)^{-1}$       \\\hline
$c2$    & $\ct$    & $0.5\,\s^{-1}$        \\\hline
$c3$    & $\cth$   & $0.00048\,\s^{-1}$    \\\hline
$c4$    & $\cf$    & $0.0045\,\s^{-1}$      \\\hline
$ki3a$  & $\kitha$ & $0.00067\,\s^{-1}$       \\\hline
$ke3a$  & $\ketha$ & $3.35\E{-4}\,\s^{-1}$     \\\hline
\end{tabular}
\begin{tabular}{\fmt}
\hline\multicolumn{3}{|c|}{\textbf{Parameters}} \\\hline
$ki1$   & $\kio$   & $0.0026\,\s^{-1}$      \\\hline
$ke1$   & $\keo$   & $0.000052\,\s^{-1}$    \\\hline
$kc1a$  & $\kcoa$  & $0.074\,\s^{-1}$  \\\hline
$kc2a$  & $\kcta$  & $0.37\,\s^{-1}$  \\\hline
$kt2a$  & $\ktta$  & $0.1\,\s^{-1}$        \\\hline
$kp$    & $\kp$    & $0.0006\,\s^{-1}$      \\\hline
$kbA20$ & $\kbA$   & $0.0018\,\uM$   \\\hline
$ka$    & $\ka$    & $0.004\,\s^{-1}$       \\\hline
$ki$    & $\ki$    & $0.003\,\s^{-1}$      \\\hline
$TR$    & $\TR$    & $1/0$           \\\hline
$k$     & $k$      & $0.065\,\uM$       \\\hline
$h$     & $h$      & $2$           \\\hline
\blank \\\hline
\blank \\\hline
\end{tabular}
\end{minipage}}
\end{table}

The ``two-feedback'' model of the \NFKB\ system
presented in \citet{Ashall-etal-2009} is our starting point. It is a system of 14
ordinary differential equations representing \NFKB\ and the \IKBA\ and A20
negative feedbacks (\fig{Fig1}).  We use brief notations
for its variables and parameters as given in \tab{1}. We pursued derivation of a minimal model with respect to 
a representative solution obtained for initial conditions as described in~\tab{1} and $\TR=1$. In a biological 
 context this corresponds to a continuous stimulation of the system with a high 
 dose of \TNFA\ \citep{Ashall-etal-2009}.

Before employing the reduction algorithm we
endeavoured to simplify the system by elementary means~(similarly to~\cite{Wang-etal-2012}). Conservation of cellular IKK reads $\IKKn+\IKKa+\IKKi=\IK=\const$, which allows us to eliminate $\IKKi$ via the substitution $\IKKi=\IK-\IKKn-\IKKa$. Similarly, conservation of \NFKB\ in all its five forms reads $\NFkB+\IkBaNFkB+\pIkBaNFkB+\frac1\kv(\nNFkB+\nIkBaNFkB)=\NF=\const$, which we use to eliminate $d$. Further, we observed that variable $\pIkBa$ is ``decoupled'': 
it is only present in its own equation, and the dynamics of other variables do not
depend on it. So it can be removed from the analysis, as the solution
for it, if necessary, can be obtained post factum by integration of the solution of the remaining 
system. Finally, for this representative solution we have observed that some of the terms in 
the equation consistently remain so small that their elimination does not visibly
change the solution. This involved elimination of variable $\nIkBaNFkB$, leaving a system of 10 equations, which we shall refer to as the Simplified Model (SM): 
\begin{subequations}                              \label{sm}
  \begin{align}
    \Df{\NFkB}{t} &= \ktta\pIkBaNFkB -\kaoa\IkBa\NFkB-\kio\NFkB+\keo\nNFkB
    \displaybreak[2]\\
    \Df{\IkBa}{t} &= -\kaoa\IkBa\NFkB+\cta\tIkBa-\cfa\IkBa-\kitha\IkBa+\ketha\nIkBa-\kcoa\IKKa\IkBa
    \displaybreak[2]\\
    \Df{\nNFkB}{t} &= \kio\kv\NFkB-\kaoa\nIkBa\nNFkB-\keo\kv\nNFkB
    \displaybreak[2]\\
    \Df{\nIkBa}{t} &= \kitha\kv\IkBa-\kaoa\nIkBa\nNFkB-\cfa\nIkBa-\ketha\kv\nIkBa
    \displaybreak[2]\\
    \Df{\tIkBa}{t} &= \coa\frac{\nNFkB^{h}}{\nNFkB^{h}+k^{h}}-\ctha\tIkBa
    \displaybreak[2]\\
    \Df{\IKKn}{t} &= \kp\frac{\kbA}{\kbA+\TR\A}\left(\IK-\IKKn\right)-\TR\ka\IKKn
    \displaybreak[2]\\
    \Df{\IKKa}{t} &= \TR\ka\IKKn-\ki\IKKa
    \displaybreak[2]\\
    \Df{\tA}{t} &= \co\frac{\nNFkB^{h}}{\nNFkB^{h}+k^{h}}-\cth\tA
    \displaybreak[2]\\
    \Df{\A}{t} &= \ct\tA-\cf\A
    \displaybreak[2]\\
    \Df{\pIkBaNFkB}{t} &= \kcta\IKKa\left(\NF-\NFkB-\frac{\nNFkB}{\kv}\right)-\ktta\pIkBaNFkB
  \end{align}
\end{subequations}
The solution of~\eq{sm} is very close to that of the original model
(see~\tab{2} and Appendix~\ref{app:SM}), and marks the starting point of the
reduction procedure.  We apply the reduction algorithm iteratively,
eliminating a sequence of fast variables and employing different orders of
approximation for them.  To keep track of these, we introduce a
nomenclature for the reduced models. The model variants are named
according to the variables that have been removed, each with a
subscript showing if a zeroth- or first-order QSSA has been used, $0$
or $1$ respectively. For example, the first variable eliminated is
$\pIkBaNFkB$, therefore the model with this variable replaced with a
zeroth-order QSSA is titled $\pIkBaNFkB_0$ and the same with a
first-order QSSA is titled $\pIkBaNFkB_1$. A model where the variables
$\pIkBaNFkB$ and $\NFkB$ have been replaced in turn with their zeroth-
and first-order QSSAs respectively, will be denoted as
$\pIkBaNFkB_0\NFkB_1$, etc. Below, we concentrate on the key points of
the reduction sequence.
\begin{figure}[tbp]
\centerline{\includegraphics[width=1\textwidth]{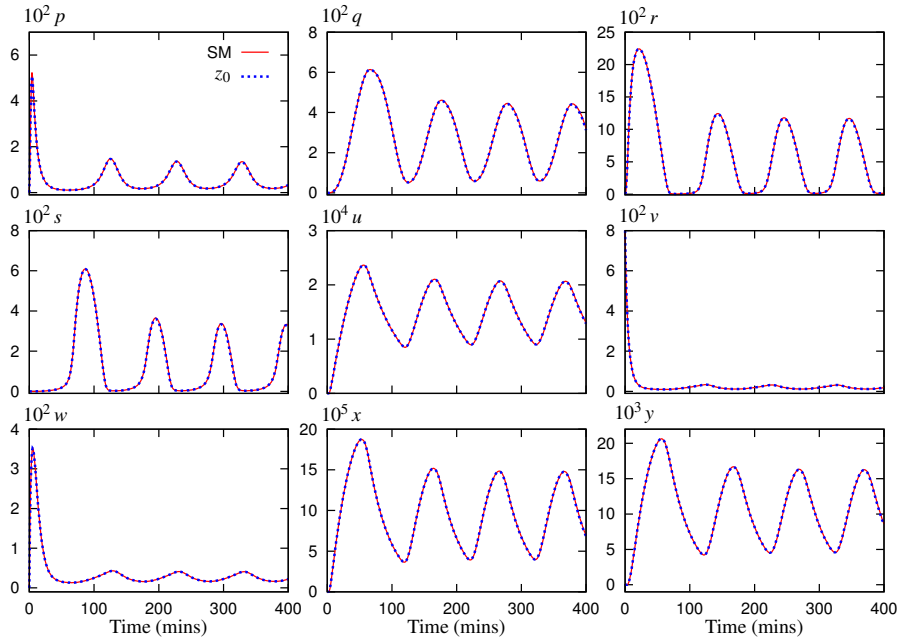}}
\caption{
Components of the representative solution for the 10-variable
Simplified Model (SM, solid lines) and reduced 9-variable model
$\pIkBaNFkB_0$ (dashed lines).  The lines visually coincide in all
cases, indicating that zeroth-order approximation is sufficient for
$\pIkBaNFkB$.
}
\label{fig:Fig3}
\end{figure}

\Fig{Fig2} shows the speed coefficients calculated for the Simplified
Model. It identifies $\pIkBaNFkB$ to be the fastest and thus eliminated first. Application of the
method, using zeroth-order approximation, results in a 9-variable
model $\pIkBaNFkB_0$ with comparable solution to this of the Simplified
Model (\fig{Fig3}).

\begin{figure}[tbp]
\centerline{\includegraphics[width=1\textwidth]{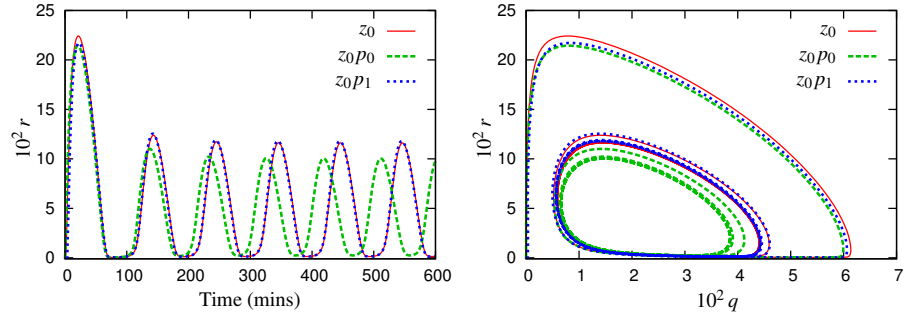}}
\caption{\
Comparison of the representative solution for the 9-variable model
$z_0$ (solid lines) and its two 8-variable reductions, with the
zeroth-order (dashed lines) and the first-order (dotted lines)
approximations for $\NFkB$.  Left panel shows a solution for the
variable nuclear NF-$\kappa$B, $\nNFkB$, right panel shows a phase
plane for the variables $\IkBa$ and $\nNFkB$. Use of the first-order
approximation gives a marked improvement in accuracy of the reduced
model.
}
\label{fig:Fig4}
\end{figure}

Addition of a first-order correction to some of the QSSAs improved the
model fit in comparison to respective predecessors. \Fig{Fig4} shows that  a
first-order correction in the variable $p$ markedly improved the
accuracy of the 8-variable reduced model. However, addition of
these corrections can also increase the algebraic complexity of the
system and it must be considered whether the improvement of the model 
outweighs the added complexity.

\begin{figure}[tbp]
\centerline{\includegraphics[width=0.7\textwidth]{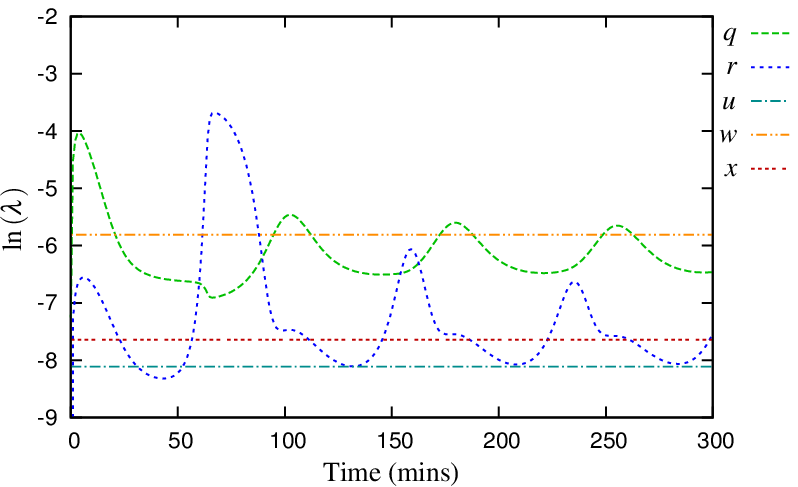}}
\caption{
Semi-logarithmic plot of speed coefficients for dynamic variables of
the $z_0p_{0}y_{0}v_{0}s_{0}$ model.  The variable $\IKKa$ has the
largest minimum compared to other variables, identifying it as the
most appropriate candidate for the next elimination.
}
\label{fig:Fig5}
\end{figure}

As the reduction progressed, there was an increasing overlap in the
ranges of the speed coefficients, and we had to apply the ``the
  fastest at its slowest'' heuristic rule.
For example  in
\fig{Fig5}, this rule identifies the variable $\IKKa$ for elimination during reduction
to the 4-variable model, even though two
other variables, $\nNFkB$ and $\IkBa$, are at times faster.

\begin{table}[tbp]
  \caption{
    Key features of \NFKB\ oscillations for each of the model
    variants. Fold change in period and amplitude was calculated relative
    to the period and amplitude of the original model in
    \citep{Ashall-etal-2009}. MSE was calculated after the models had been
    scaled to have the same period.
  }
\label{tab:2}
\begin{tabular}{|p{2.1cm}|p{1.8cm}|p{1.3cm}|p{1.4cm}|p{1.4cm}|p{1.1cm} |}
\hline
\textbf{Model} & Variable removed at this stage & Period (mins) &
  Fold change in period & Fold change in amplitude & Shape MSE   $\times10^5$
  \\ \hline\hline
\textbf{Original Model}       & N/A      & 99.5  & 1    & 1    & N/A  \\ \hline
\textbf{Simplified Model}     & $a,b,c,d$& 100.5 & 1.01 & 1.06 & 1.24 \\ \hline
\textbf{$z_0$}            & $\pIkBaNFkB$ & 100.0 & 1.01 & 1.06 & 1.05 \\ \hline
\textbf{$z_0p_0$}             & $\NFkB$  & 92.5  & 0.93 & 0.92 & 2.47 \\ \hline
\textbf{$z_0p_0y_0$}          & $\A$     & 85.5  & 0.86 & 0.83 & 9.42 \\ \hline
\textbf{$z_0p_0y_0v_0$}       & $\IKKn$  & 77.7  & 0.78 & 0.65 & 32.4 \\ \hline
\textbf{$z_0p_0y_0v_0s_0$}    & $\nIkBa$ & 75.0  & 0.75 & 0.62 & 38.6 \\ \hline
\textbf{$z_0p_0y_0v_0s_0w_1$} & $\IKKa$  & 73.8  & 0.74 & 0.71 & 22.3 \\ \hline
\textbf{$z_0p_1y_0v_0s_0w_1$} & As above & 80.3  & 0.80 & 0.90 & 5.28 \\ \hline
\textbf{$z_0p_1y_1v_1s_1w_1$} & As above & 86.6  & 0.87 & 1.01 & 6.31 \\ \hline
\end{tabular}
\end{table}

\begin{figure}[tbp]
\centerline{\includegraphics[width=1\textwidth]{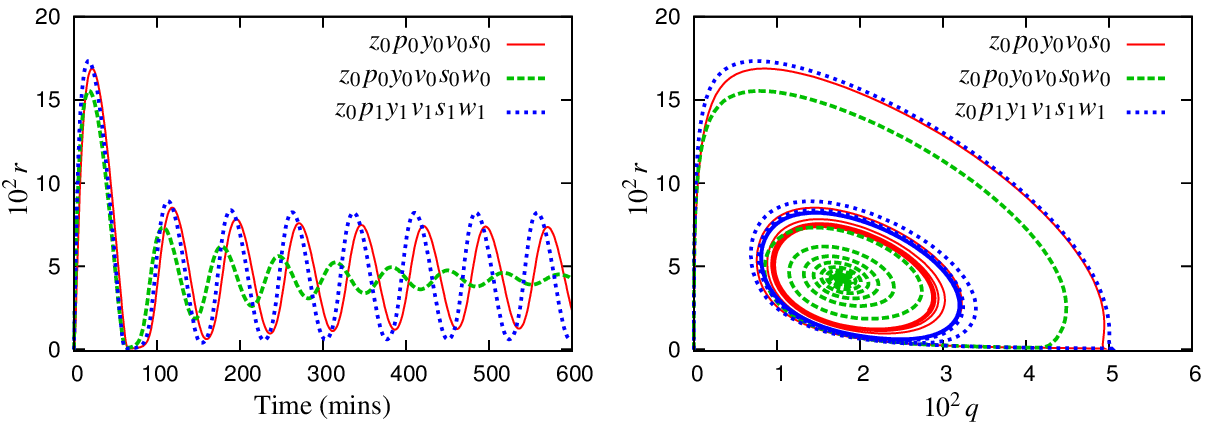}}
\caption{
Comparison of the representative solution for the 5-variable model
$z_0p_0y_0v_0s_0$ (solid lines) and its two possible 4-variable
reductions, with the zeroth-order (dashed lines) and the first-order
(dotted lines) approximations for $\IKKa$. Use of the first-order
approximation not only improved the accuracy of the 4-variable model,
but also maintained a stable limit cycle.
}
\label{fig:Fig6}
\end{figure}

Successive cycles of the algorithm were applied to ultimately
reduce this system to four differential equations. The method maintained
the important qualitative features of the system, such as the limit
cycle. However, through each stage of the reduction, the resulting
limit cycle had a slightly reduced period and amplitude
(\tab{2}). Using only the zeroth-order QSSAs was sufficient to
reduce the model to five ODEs ($z_0p_0y_0v_0s_0$), while maintaining the limit cycle. In order to reduce
the system further, the use of a first-order
QSSA was necessary (\fig{Fig6}). A
suitable zeroth- or first-order QSSA could not be calculated to reduce the
model beyond this, and therefore the model $z_0p_0y_0v_0s_0w_1$
of four differential equations was chosen as
the end point of this analysis. This minimal model is given by ~\eq{mm}, where $A= \TR\ka\kp\kbA\cf\NF$ 
and $B(\tA)=\kp\kbA\cf+\TR\ka\kbA\cf+\TR^{2}\ka\ct\ki\tA$.
\begin{subequations}                              \label{mm}
  \begin{align}
\Df{\IkBa}{t} & = 
  -\kaoa\IkBa\bar{p}(\IkBa,\nNFkB,\tA)
  +\cta\tIkBa
  -\cfa\IkBa
  -\kitha\IkBa
  +\ketha\bar{s}(\IkBa,\nNFkB)
  -\kcoa\bar{w}(\nNFkB,\tA)\IkBa
  \label{mm1} \displaybreak[2]\\
\Df{\nNFkB}{t} &= 
  \kio\kv\bar{p}(\IkBa,\nNFkB,\tA)
  -\kaoa\bar{s}(\IkBa,\nNFkB)\nNFkB
  -\keo\kv\nNFkB
  \label{mm2} \displaybreak[2]\\
\Df{\tIkBa}{t} & =
  \coa\frac{\nNFkB^{h}}{\nNFkB^{h}+k^{h}}
  -\ctha\tIkBa
  \label{mm3} \displaybreak[2]\\
\Df{\tA}{t} & =
  \co\frac{\nNFkB^{h}}{\nNFkB^{h}+k^{h}}
  -\cth\tA
  \label{mm4} \displaybreak[2]\\
  \bar{w}(\nNFkB,\tA) &= 
 \frac{
  A
 }{
 B(\tA) }
  + \frac{
  \TR^{2}\ka\ct A\left(\frac{\co\nNFkB^{h}}{\nNFkB^{h}+k^{h}}-\cth\tA\right)
 }{
B(\tA)^{2}
 }
  \label{mm5} \displaybreak[2]\\
\bar{s}(\IkBa,\nNFkB) &=
  \frac{\kv\kitha\IkBa}{\kaoa\nNFkB+\kv\ketha+\cfa}
  \label{mm6} \displaybreak[2]\\
\bar{p}(\IkBa,\nNFkB,\tA) &=\frac{
    \keo\nNFkB+\kcta\bar{w}(\nNFkB,\tA)\left(\NF-\frac{\nNFkB}{\kv}\right)
  }{
    \kaoa\IkBa+\kio+\kcta\bar{w}(\nNFkB,\tA)
  }
  \label{mm7}
\end{align}
\end{subequations}

\begin{figure}[tbp]
\centerline{\includegraphics[width=1\textwidth]{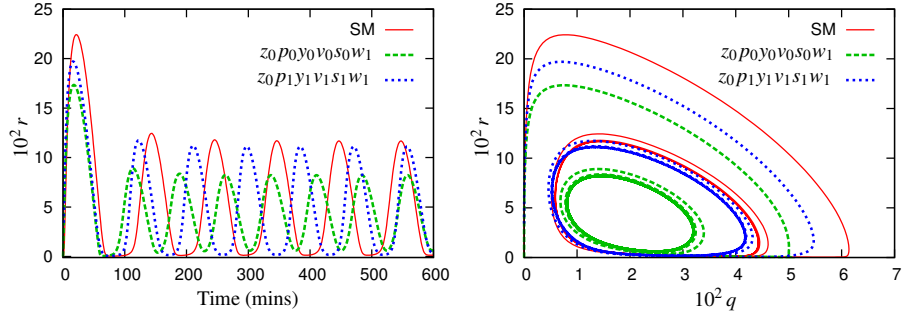}}
\caption{
Comparison of the representative solution for the Simplified Model
(solid lines) and the two four-variable reductions, the cruder
$z_0p_0y_0v_0s_0w_1$ model (dashed lines) and the more accurate
$z_0p_1y_1v_1s_1w_1$ model (dotted lines).
}
\label{fig:Fig7}
\end{figure}

It was possible to add first-order corrections to all of the dynamic variables
during the model reduction, producing a minimal model $z_1p_1y_1v_1s_1w_1$
with a far improved fit in comparison to the original. However, the
$z_0$ approximation was so accurate that $z_1$ did not make a
noticeable improvement. 
\Fig{Fig7} shows comparison of the ``simplest'' and ``the
most accurate'' 4-variable models to the original 10-variable one (the $z_0p_1y_1v_1s_1w_1$ model is presented in appendix A). 

\Fig{Fig8} shows how some of the dynamic properties of the model change
through the reduction process. It represents the steady state solution
and continuation for the variable $\nNFkB$ as the parameter $\TR$ is
varied \citep{Doedel,Ermentrout}, showing the effect of altering the \TNFA\ 
dose \citep{Turner-2010}. In the original model, there is a supercritical Hopf
bifurcation (HB) at $\TR=0.36$ above which the
limit cycle is observed. Successive elimination of the fastest
variables causes the HB point to move up, closer to the value $\TR=1$, 
which corresponds to a saturating dose of \TNFA. Reduction from five to four
differential equations using zeroth-order QSSA for $\IKKa$ would move the HB point further to
the right (Hopf bifurcation at $\TR=3.105$). \Fig{Fig8} also demonstrates that use of the
first-order correction terms dramatically reduces the loss in limit
cycle amplitude and change in the location of the HB point.

\begin{figure}[tbp]
\centerline{\includegraphics[width=1\textwidth]{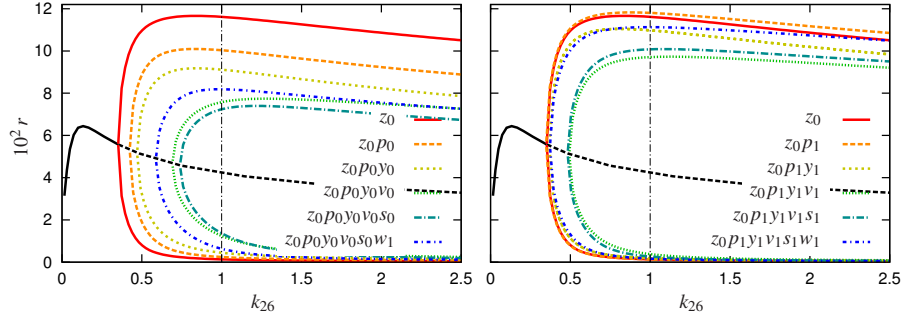}}
\caption{
Bifurcation analyses of reduced models with respect to the parameter
$\TR$, representing the dose of \TNFA\ stimulation. Branches of the
solution in colour represent minimal and maximal values of the limit
cycle. Solid and dashed black lines correspond to stable and unstable
equilibria, respectively.
}
\label{fig:Fig8}
\end{figure}

\section{Model reduction with respect to pulsed \TNFA\ input }
\label{sec:Ashall-pulse}

Previously, we derived models with respect a solution that corresponded to a
constant value of the \TNFA\ input, $\TR\equiv1$. The universality of such models depends on how representative that solution actually
is. In this subsection we give an example where a different selection
of the representative solution leads to a different reduced model.

We now consider another
experimentally relevant case, where the \TNFA\ input is varied: $\TR=0$
except for 5-minute pulses of $\TR=1$ delivered every
100~minutes. Under such stimulation, the system exhibits pulses of the
nuclear \NFKB\ entrained to the input
frequency~\citep{Ashall-etal-2009}.
Despite the same 100 minute period, these pulses are markedly different than oscillations induced with the continuous \TNFA\ input. The Simplified Model
reproduces this property, see \fig{Fig2} vs \fig{Fig9} . However, the 6-variable variant, $z_0p_0y_0v_0$ (see appendix B for equations), does not respond with a full-size nuclear \NFKB\ translocation to each pulse, and the solution is of a double period, ~\fig{Fig9}.

\begin{figure}[tbp]
\centerline{\includegraphics[width=1\textwidth]{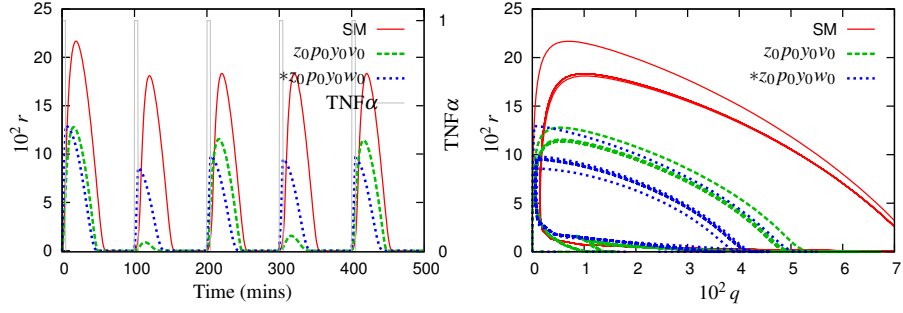}}
\caption{
Models' response to a pulsed \TNFA\ input. Shown are solution of the
Simplified Model (SM, solid line), the 6-variable model
$z_{0}p_{0}y_{0}v_{0}$ (dashed line), and the alternative 6-variable
model $\ast z_{0}p_{0}y_{0}v_{0}$ ~\eq{pmm} (dotted line). The \TNFA\
input is varied: $\TR=0$ except for 5-minute pulses of $\TR=1$
delivered every 100~minutes (shown in grey lines on the left panel).
}
\label{fig:Fig9}
\end{figure}
\begin{figure}[tbp]
\centerline{\includegraphics[width=1\textwidth]{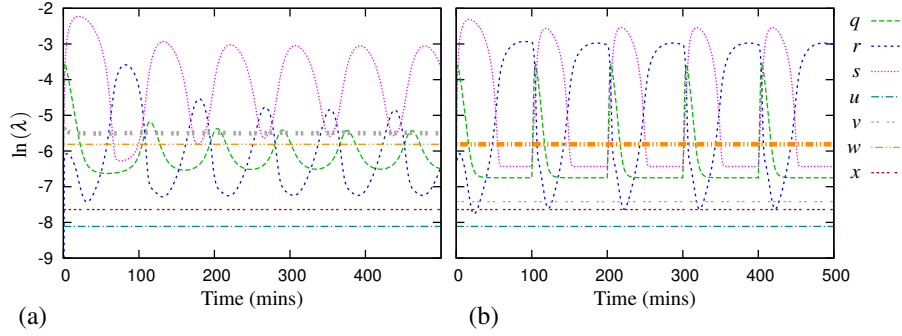}}
\caption{
Comparison of the speed coefficients for the $z_0p_0y_0$ calculated
with respect to different solutions.  (a) Constant input
($\TR\equiv1$). (b) Pulsed input; $\TR=0$ except for 5-minute pulses
of $\TR=1$ delivered every 100~minutes. Depicted in bold are the
fourth fastest variables: $v$ in (a) and $w$ in (b).
}
\label{fig:Fig10}
\end{figure}

We therefore developed an alternative minimal model, choosing the
periodically entrained solution as the representative one.  For the periodically entrained
solution, the hierarchy of speeds of the variables associated with the IKK module is different from
the $\TR\equiv1$ case. Specifically, the first three fastest variables
are $z$, $p$ and $y$ as before. However,
when choosing the 4th variable for elimination, the neutral form of IKK, $\IKKn$, becomes one of the slowest, and the
algorithm identified the active IKK, $\IKKa$, for approximation (\fig{Fig10}). In the continuous case, $\IKKn$ and $\IKKa$ were the first and second fastest variables, respectively (\fig{Fig10}). Ultimately, application of the algorithm with respect to the pulsed input resulted in a different model, which showed a much better agreement with the 
SM and did not display a period doubling (\fig{Fig9}). This alternative
6-variable, ($\ast z_0p_0y_0w_0$) model is given by:
\begin{subequations}                              \label{pmm}
  \begin{align}
    \Df{\IkBa}{t} &= -\kaoa\IkBa\bar{p}+\cta\tIkBa-\cfa\IkBa-\kitha\IkBa+\ketha\nIkBa-\kcoa\bar{w}\IkBa
    \displaybreak[2]\\
    \Df{\nNFkB}{t} &= \kio\kv\bar{p}-\kaoa\nIkBa\nNFkB-\keo\kv\nNFkB
    \displaybreak[2]\\
    \Df{\nIkBa}{t} &= \kitha\kv\bar{p}-\kaoa\nIkBa\nNFkB-\cfa\nIkBa-\ketha\kv\nIkBa
    \displaybreak[2]\\
    \Df{\tIkBa}{t} &= \coa\frac{\nNFkB^{h}}{\nNFkB^{h}+k^{h}}-\ctha\tIkBa
    \displaybreak[2]\\
    \Df{\IKKn}{t} &= \kp\frac{\kbA}{\kbA+\TR\A}\left(\IK-\IKKn\right)-\TR\ka\IKKn
    \displaybreak[2]\\
    \Df{\tA}{t} &= \co\frac{\nNFkB^{h}}{\nNFkB^{h}+k^{h}}-\cth\tA
    \displaybreak[2]\\
    \bar{p} &= \frac{\keo\nNFkB\kv+\kcta\bar{w}\NF\kv-\kcta\bar{w}\nNFkB}{\kv\left(\kaoa\IkBa+\kio+\kcta\bar{w}\right)}
    \displaybreak[2]\\
    \bar{y} &= \frac{\ct\tA}{\cf}
     \displaybreak[2]\\
   \bar{w} &= \frac{\TR\ka\IKKn}{\ki}
  \end{align}
\end{subequations}
The difference in the $\IKKn$ speed for alternative \TNFA\ stimulation can be easily understood by
analysing the dynamic equation for $\IKKn$.  The last term in its
right-hand side, $-\TR\ka\IKKn$, directly contributes towards decay of
$\IKKn$, but only when $\TR$ is switched on. So when $\TR$ is off, the
$\IKKn$ variable is much slower and its adiabatic elimination is not
justified.

\section{Application of speed coefficients method to Krishna model}
\label{sec:Krishna}

Here, we compare the behaviour and properties of
two reduced models of Krishna's full 6-variable model for \NFKB\
signalling dynamics~\citep{Krishna}, one obtained by combination of
coarse graining and numerical observations, and the other obtained
using our new method of speed coefficients. In this analysis, we
demonstrate better agreement with the full model achieved using our
algorithmic approach.

\begin{figure}[tbp]
\centerline{\includegraphics[width=1\textwidth]{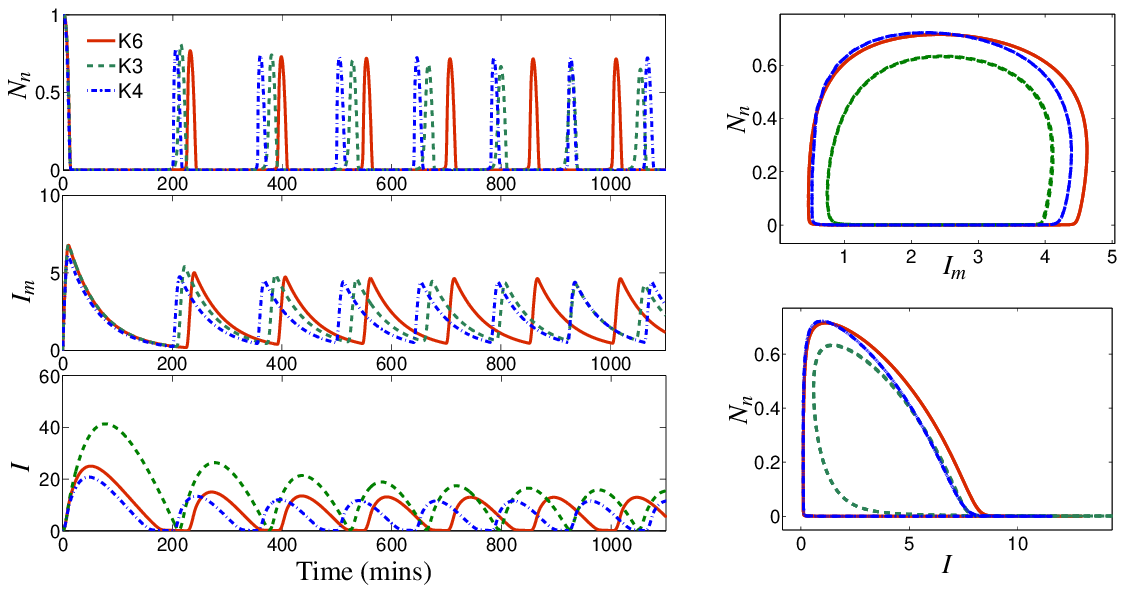}}
\caption{
Analysis of alternatively reduced models of the \NFKB\
system. (Left-hand panel) Time courses for the 3-variable reduced
model (K3) and its 6-variable predecessor developed in~\citet{Krishna}
(K6), together with a new 4-variable reduced model obtained using the
speed coefficient method (K4). Variables $N_{n},\;I_{m},\;I$ represent
nuclear \NFKB, \IKBA\ protein and \IKBA\ mRNA
respectively. (Right-hand panel) Corresponding phase portraits for the
limit cycles that the respective systems approach.
}
\label{fig:tc_and_pp_643}
\end{figure}

\begin{figure}[tbp]
\centerline{\includegraphics[width=1\textwidth]{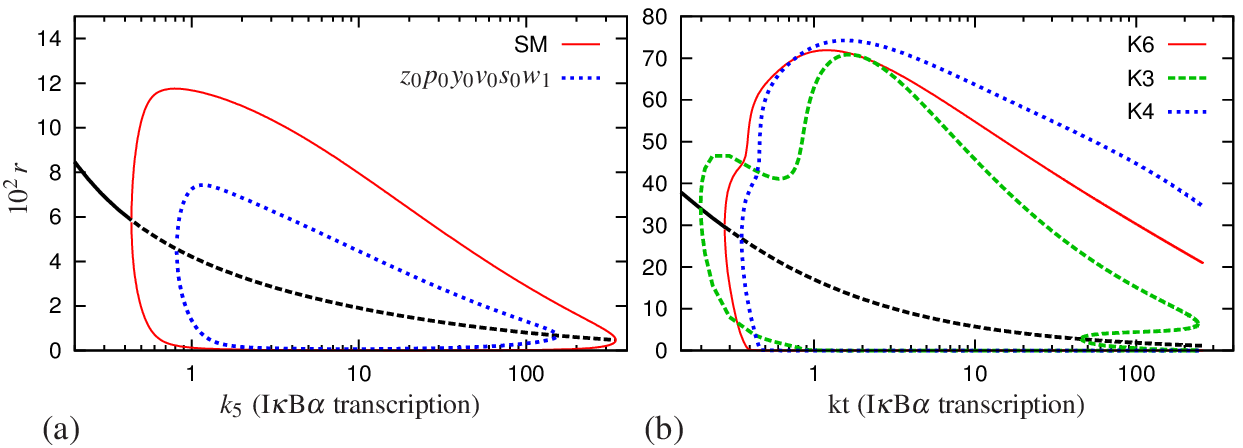}}
\caption{
Analysis of alternatively reduced models of the \NFKB\ system. Shown
are bifurcation diagrams with respect to the rate of the \IKBA\
transcription. Simulations performed for a continuous \TNFA\ input.
(a) The minimal $z_0p_0y_0v_0s_0w_1$ model developed herein ~\eq{mm}
in comparison to the SM. The \IKBA\ transcription rate was normalised
to its nominal value (as in \tab{1}). (b) The 3-variable reduced model
(K3) and its 6-variable predecessor developed in~\citet{Krishna} (K6),
together with a new 4-variable reduced model obtained using the speed
coefficient method (K4). Over the range of $kt$ shown, the new reduced
model gives better qualitative agreement with the full model, and does
not introduce the subcritical Hopf bifurcation seen in K3.
}
\label{fig:Fig11}
\end{figure}

Firstly, in \fig{tc_and_pp_643}, we show time courses for oscillatory
solutions for variables representing nuclear \NFKB, \IKBA\ protein and
\IKBA\ mRNA in three models, namely Krishna's full model (K6),
Krishna's 3-variable minimal model (K3), and a new 4-variable reduced
model given by our speed coefficient algorithm (K4) (see
Appendix~\ref{app:krishna} for the systems of equations). We note
that, while neither of the reduced models matches the full model in
period, the oscillation amplitudes of the three variables show
reasonable agreement, with our new reduced model (K4) more closely
agreeing with the full model. Also, the K4 \IKBA\ protein profile
shape shows better agreement with K6 than K3 does, with $I$ flattening
out in its troughs. Summary phase portraits clearly show that K4's
limit cycles more closely agree with K6 than K3 does.

In \fig{Fig11} we compare bifurcation diagrams (with respect to the
rate of \IKBA\ transcription) for reduced models with their
corresponding full models, for both our Simplified Model and the
Krishna model. For the Krishna model, we further compare the Krishna
minimal model (K3)and our new reduced model (K4). The reduced model
resulting from the speed coefficient method applied to the Simplified
Model (SM) gives a bifurcation diagram in qualitative agreement with
that for the corresponding full model over the range of $k_{5}$ shown
(\fig{Fig11}(a)). Also, the reduced model resulting from our method
applied to the Krishna model (see Appendix~\ref{app:krishna}) gives
qualitative agreement with the full Krishna model
(\fig{Fig11}(b)). This is a marked improvement over the Krishna
minimal model, which demonstrates features that are not present in the
corresponding full model. These include variation of the limit cycle
amplitude for values of the \IKBA\ transcription around 1, and a
subcritical Hopf bifurcation at around $kt=50$, with unstable limit
cycles and hysteresis for the values between 50 and 240. On the
contrary, our minimal model preserves the properties of the full model
at least at the qualitative level, even for the values of the
parameter very different from the one corresponding to the
representative solution.

We conclude that application of our method of speed coefficients can
produce a reduced model of comparable dimensionality while better
preserving the dynamic properties of the original system than other
existing techniques.

\section{Discussion}

A key problem in computational and systems biology is to understand
how dynamical properties of a system arise via the
underlying biochemical networks. However, as these networks involve
many components this task becomes analytically intractable
and computationally challenging. In this manuscript we present a
clearly defined and accessible QSSA algorithm for reduction of such
biochemical reaction systems.  The method proposed relies on the
derivation of speed coefficients to rank system variables according to 
how quickly they approach their momentary-steady state. This enables 
a systematic method for selection of variables for steady-state approximation 
at each step of the algorithm. 

We used the method to derive a minimal models of the \NFKB\ signalling
network, a key regulator of the immune 
response~\citep{Hayden}. 
Single cell time-lapse analyses showed that the \NFKB\ system exhibits
oscillatory dynamics in response to cytokine
stimulation~\citep{Nelson,Turner-2010,Tay-etal-2010}. It has been
shown that the frequency of those oscillations may govern downstream
gene expression and therefore be the key functional output of the
system \citep{Ashall-etal-2009,Sung-2009,Tay-etal-2010}. The ability
to control the \NFKB\ dynamics may therefore provide novel ways to
treat inflammatory disease~\citep{Paszek-etal-2010}.

\NFKB\ dynamics are generated via a complex network involving several
negative feedback genes, such as A20 and \IKBA\
\citep{Hoffmann_circuitry}. Many mathematical models have been
developed to recapitulate existing experimental data by quite complex
biochemical networks involving up to 30 dynamic variables and 100
parameters with varying degrees of accuracy
\citep{Hoffmann_2002,Lipniacki-etal-2004,Radulescu-etal-2008}.
Sensitivity analyses have then demonstrated that several parameters
related to feedback regulation and IKK activation are responsible for
generation of the oscillatory dynamics
\citep{Ihekwaba,Ihekwaba-2005,Sung-2009}. An interesting extension of
the sensitivity analysis method was proposed by
\citet{Jacobsen-Cedersund-2008} who considered sensitivity with
respect not just parameter perturbations but to variations of the
network structure, e.g. introduction of delays in the network
connections. Model reduction discussed in our paper provides an
alternative avenue to extract core network components. Indeed,
minimal models by Krishna \etal\ and Radulescu \etal\ 
demonstrated that part of this complex system in response to
continuous cytokine stimulation may be reduced to three dynamical
variables describing the nuclear \NFKB\, and \IKBA\ mRNA and
protein~\citep{Krishna}. Here, we apply our method of speed
coefficients to systematically reduce a 2-feedback model of the \NFKB\
system by Ashall et al~\citep{Ashall-etal-2009}.

Starting from a 14-variable model, we succeeded in closely
representing dynamics of the \NFKB\ network in response to constant
\TNFA\ input by a set of four variables~\eq{mm}. The minimal model
included the nuclear \NFKB\ and its cytoplasmic inhibitor
I$\kappa$B$\alpha$, as well as two negative feedback loops represented
by \IKBA\ and A20 transcripts. The latter variables were consistently
ranked the slowest during successive reduction steps (\fig{Fig2} and
\fig{Fig5}), and in fact their subsequent QSSA resulted in the loss of
oscillations. This suggested that the timescale of transcription
relative to other processes generates the key delayed negative
feedback motif that drives oscillations in the system
\citep{Novak-2008}. While reducing the model, we observed that the
period as well as the amplitude of oscillations was decreased with
each reduction (\tab{2}). Replacing those variables with the
respective QSSAs decreased the effective delay time in the system, and
thus reduced the system's propensity for oscillations. This effect was
reverted by using first-order QSSA for some of the eliminated
variables, namely cytoplasmic NF-$\kappa$B, nuclear \IKBA\ and the
active form of IKK kinase. A more accurate representation of those
variables is thus important to faithfully represent \NFKB\ dynamics
(\fig{Fig6} - \fig{Fig8}). 
Our analysis is in agreement with results of Radulescu \etal\ who,
using quasi-stationarity arguments, obtained a series of reduced
models and eventually arrived at the 5 variable minimal model.  While
starting from a different two-feedback IkBa and A20 model
\citep{Lipniacki-etal-2004} than the one considered here, Radulescu
\etal\ showed similar requirements for both feedbacks to
maintain oscillatory dynamics.

A model derived with respect to a specific solution is not necessarily
able to reproduce the same breadth of responses as its
forebear. However, by applying the algorithm with respect to a
different solution one might try to potentially extract other key
features of the system. Here, we demonstrated that the reduction of
the model with respect to a pulsed and continuous \TNFA\ input
resulted in a different order of elimination of the variables and
ultimately a different minimal models (\fig{Fig9}). The differences
unravelled specific components of the IKK module responsible for
\NFKB\ dynamics in response to different stimulus. With a pulsed input
the amplitude of the subsequent peaks is determined by the
``refractory period'', i.e. the time it takes for the active IKK to
return to its neutral state. This requires a very accurate temporal
representation of the neutral form of IKK, $\IKKn$, in the
model. However, in response to continuous \TNFA\ input, both
IKK-related variables became less important, and their steady-state
approximation is sufficient to support the limit cycle. This analysis
therefore begins to unravel how components of the KK signalling module
could differentially encode temporal inflammatory signals.

In order to demonstrate a more general applicability of our method, we
have employed the speed coefficient algorithm to derive a new reduced
model of the Krishna model~\citep{Krishna}. The comparison with
minimal Krishna et al model showed that both models perform similarly
in terms of time courses and phase portraits (fig. 11). However,
analysis of bifurcation diagrams showed that our algorithmic approach
better preserved dynamical properties of the system (fig.12). In fact,
the Krishna minimal model demonstrates features such as
unstable limit cycle and hysteresis that are not present in the
corresponding full model. 
Recently, Kourdis \etal\ used CSP algorithm to asymptotically analyse
the dynamics of the Krishna \etal\ model. In agreement with our
approach, their analysis identified similar fast/slow time scale
variables that are essential to recapitulate limit cycle behaviour of
the system.
This analysis, in addition to our discussion
of the Simplified Model, certainly suggests that our method has
further potential as a viable technique for the reduction of
biochemical network dynamic models.

Our objective here was to present and implement a new model reduction
technique that without relying on prior biological insights, would
preserve characteristics of the original model's numerical
solutions. This method thus belongs to a class of reduction methods
that are algorithmic rather than biologically or biochemically
intuitive, and as such should be applicable to complex biochemical
models where the most important network sub-structures underlying the
observed dynamical behaviour are not necessarily apparent.
Similarly to other approximation methods, there is a
trade-off between simplicity and accuracy of the end-point
models. Even if errors introduced by one reduction step are small, for
many steps they can accumulate. The approximations can be improved by
using higher-order asymptotics, which increases algebraic complexity
of the resulting reduced model but retains the dimensionality. We
believe that in practically interesting cases, the increased algebraic
complexity can be overcome by appropriate approximation of the
functions in the resulting models. Another way to improve the accuracy
of reduced models is to adjust parameters to match the solutions of
the full model; a semi-empirical model resulting from such adjustment
would still have an advantage over a fully empirical model in that at
least its structure is not arbitrarily postulated. In addition to a
lower dimensionality, the reduced problems are less stiff, as by
definition, the variables with fastest characteristic timescales are
eliminated first. The reduced dimensionality and stiffness allow, in
principle, more efficient computations which may be important, e.g.,
for large scale models including interaction of many cells. Last but
not least, systems of lower dimensionality are more amenable for
qualitative study and intuitive understanding.
	
\subsection*{Acknowledgements}

\noindent The authors are grateful to Dr I.V. Biktasheva and Dr R.N. Bearon at the 
University of Liverpool for stimulating discussions and critical reading.

\bibliographystyle{spbasic}      

\begin{thebibliography}{46}
\providecommand{\natexlab}[1]{#1}
\providecommand{\url}[1]{{#1}}
\providecommand{\urlprefix}{URL }
\expandafter\ifx\csname urlstyle\endcsname\relax
  \providecommand{\doi}[1]{DOI~\discretionary{}{}{}#1}\else
  \providecommand{\doi}{DOI~\discretionary{}{}{}\begingroup
  \urlstyle{rm}\Url}\fi
\providecommand{\eprint}[2][]{\url{#2}}

\bibitem[{Anosov(1960)}]{Anosov-1960}
Anosov DV (1960) Limit cycles of systems of differential equations with small
  parameters in the highest derivatives. Mat Sb (NS) 50(92)(3):299--334

\bibitem[{Ashall et~al(2009)Ashall, Horton, Nelson, Paszek, Harper, Sillitoe,
  Ryan, Spiller, Unitt, Broomhead, Kell, Rand, See, and
  White}]{Ashall-etal-2009}
Ashall L, Horton CA, Nelson DE, Paszek P, Harper CV, Sillitoe K, Ryan S,
  Spiller DG, Unitt JF, Broomhead DS, Kell DB, Rand DA, See V, White MR (2009)
  Pulsatile stimulation determines timing and specificity of
  nf-kappab-dependent transcription. Science 324(5924):242--246

\bibitem[{Biktashev and Suckley(2004)}]{Biktashev-Suckley-2004}
Biktashev VN, Suckley R (2004) Non-tikhonov asymptotic properties of cardiac
  excitability. Physical Review Letters 93:168,103

\bibitem[{Biktasheva et~al(2006)Biktasheva, Simitev, Suckley, and
  Biktashev}]{Biktasheva-2006}
Biktasheva IV, Simitev RD, Suckley R, Biktashev VN (2006) Asymptotic properties
  of mathematical models of excitability. Phil Trans Roy Soc London A: Math
  Phys Eng Sci 364(1842):1283--98

\bibitem[{Briggs and Haldane(1925)}]{Briggs-Haldane-1925}
Briggs GE, Haldane JBS (1925) A note on the kinetics of enzyme action. Biochem
  J 19:338--339

\bibitem[{Dan{\o} et~al(2006)Dan{\o}, Madsen, Schmidt, and
  Cedersund}]{Dano-etal-2006}
Dan{\o} S, Madsen MF, Schmidt H, Cedersund G (2006) Reduction of a biochemical
  model with preservation of its basic dynamic properties. FEBS Journal
  273(21):4862--4877

\bibitem[{Doedel et~al(2000)Doedel, Paffenroth, Champneys, Fairgrieve,
  Kuznetsov, Sandstede, and Wang}]{Doedel}
Doedel E, Paffenroth R, Champneys A, Fairgrieve T, Kuznetsov Y, Sandstede B,
  Wang X (2000) Auto2000: Continuation and bifurcation software for ordinary
  differential equations (with homcont). Technical report, California Institute
  of Technology

\bibitem[{Ermentrout(2002)}]{Ermentrout}
Ermentrout B (2002) Simulating, analysing, and animating dymaical systems: a
  guide to XPPAUT for researchers and students., Software, Environments, and
  Tools, vol~14. SIAM

\bibitem[{Fenichel(1979)}]{Fenichel-1979}
Fenichel N (1979) Geometric singular perturbation theory for ordinary
  differential equations. Journal of Differential Equations 31:53--98

\bibitem[{Gorban and Karlin(2003)}]{Gorban-Karlin-2003}
Gorban AN, Karlin IV (2003) Method of invariant manifold for chemical kinetics.
  Chemical Engineering Science 58:4751--4768

\bibitem[{Hayden and Ghosh(2008)}]{Hayden}
Hayden MS, Ghosh S (2008) Shared principles in nf-kappab signaling. Cell
  132(3):344--362

\bibitem[{Hoffmann and Baltimore(2006)}]{Hoffmann_circuitry}
Hoffmann A, Baltimore D (2006) Circuitry of nuclear factor kappab signaling.
  Immunol Rev 210:171--186

\bibitem[{Hoffmann et~al(2002)Hoffmann, Levchenko, Scott, and
  Baltimore}]{Hoffmann_2002}
Hoffmann A, Levchenko A, Scott ML, Baltimore D (2002) The ikappab-nf-kappab
  signaling module: temporal control and selective gene activation. Science
  298(5596):1241--1245

\bibitem[{Ihekwaba et~al(2004)Ihekwaba, Broomhead, Grimley, Benson, and
  Kell}]{Ihekwaba}
Ihekwaba AE, Broomhead DS, Grimley RL, Benson N, Kell DB (2004) Sensitivity
  analysis of parameters controlling oscillatory signalling in the nf-kappab
  pathway: the roles of ikk and ikappabalpha. Syst Biol (Stevenage)
  1(1):93--103

\bibitem[{Ihekwaba et~al(2005)Ihekwaba, Broomhead, Grimley, Benson, White, and
  Kell}]{Ihekwaba-2005}
Ihekwaba AE, Broomhead DS, Grimley R, Benson N, White MR, Kell DB (2005)
  Synergistic control of oscillations in the nf-kappab signalling pathway. Syst
  Biol (Stevenage) 152(3):153--160

\bibitem[{Jacobsen and Cedersund(2008)}]{Jacobsen-Cedersund-2008}
Jacobsen EW, Cedersund G (2008) Structural robustness of biochemical network
  models-with application to the oscillatory metabolism of activated
  neutrophils. IET Syst Biol 2(1):39--47, \doi{10.1049/iet-syb:20070008}

\bibitem[{Kitano(2002)}]{Kitano}
Kitano H (2002) Computational systems biology. Nature 420(6912):206--210

\bibitem[{Klonowski(1983)}]{Klonowski-1983}
Klonowski W (1983) Simplifying principles for chemical and enzyme
  reaction-kinetics. Biophysical Chemistry 18(2):73--87

\bibitem[{Kourdis et~al(2013)Kourdis, Palasantza, and
  Goussis}]{Kourdis-etal-2013}
Kourdis PD, Palasantza AG, Goussis DA (2013) Algorithmic asymptotic analysis of
  the {NF-$\kappa$B} signaling system. Computers and Mathematics with
  Applications 65(10):1516--1534

\bibitem[{Krishna et~al(2006)Krishna, Jensen, and Sneppen}]{Krishna}
Krishna S, Jensen MH, Sneppen K (2006) Minimal model of spiky oscillations in
  nf-kappab signaling. Proc Natl Acad Sci U S A 103(29):10,840--10,845

\bibitem[{Kutumova et~al(2013)Kutumova, Zinovyev, Sharipov, and
  Kolpakov}]{Kutumova-etal-2013}
Kutumova E, Zinovyev A, Sharipov R, Kolpakov F (2013) Model composition through
  model reduction: a combined model of {CD95} and {NF-$\kappa$B} signaling
  pathways. BMC Systems Biology 7(1):13, \doi{10.1186/1752-0509-7-13}

\bibitem[{Lam and Goussis(1994)}]{Lam-Goussis-1994}
Lam SH, Goussis DA (1994) The {CSP} method for simplifying kinetics.
  International Journal of Chemical Kinetics 26(4):461--486

\bibitem[{Lipniacki et~al(2004)Lipniacki, Paszek, Brasier, Luxon, and
  Kimmel}]{Lipniacki-etal-2004}
Lipniacki T, Paszek P, Brasier AR, Luxon B, Kimmel M (2004) Mathematical model
  of nf-kappab regulatory module. J Theor Biol 228(2):195--215

\bibitem[{Maas and Pope(1992)}]{Maas-Pope-1992}
Maas U, Pope SB (1992) Simplifying chemical kinetics: Intrinsic low-dimensional
  manifolds in composition space. Combustion and Flame 88:239--264

\bibitem[{Maeda et~al(1998)Maeda, Pakdaman, Nomura, Doi, and Sato}]{Maeda-1998}
Maeda Y, Pakdaman K, Nomura T, Doi S, Sato S (1998) Reduction of a model for an
  onchidium pacemaker neuron. Biological Cybernetics 78(4):265--276

\bibitem[{Mengel et~al(2012)Mengel, Krishna, Jensen, and Trusina}]{Mengel-2012}
Mengel B, Krishna S, Jensen MH, Trusina A (2012) Nested feedback loops in gene
  regulation. Physica a-Statistical Mechanics and Its Applications
  391(1--2):100--106

\bibitem[{Nelson et~al(2004)Nelson, Ihekwaba, Elliott, Johnson, Gibney,
  Foreman, Nelson, See, Horton, Spiller, Edwards, McDowell, Unitt, Sullivan,
  Grimley, Benson, Broomhead, Kell, and White}]{Nelson}
Nelson DE, Ihekwaba AE, Elliott M, Johnson JR, Gibney CA, Foreman BE, Nelson G,
  See V, Horton CA, Spiller DG, Edwards SW, McDowell HP, Unitt JF, Sullivan E,
  Grimley R, Benson N, Broomhead D, Kell DB, White MR (2004) Oscillations in
  nf-kappab signaling control the dynamics of gene expression. Science
  306(5696):704--708

\bibitem[{Novak and Tyson(2008)}]{Novak-2008}
Novak B, Tyson JJ (2008) Design principles of biochemical oscillators. Nat Rev
  Mol Cell Biol 9(12):981--991

\bibitem[{Paszek et~al(2010)Paszek, Jackson, and White}]{Paszek-etal-2010}
Paszek P, Jackson DA, White MR (2010) Oscillatory control of signalling
  molecules. Current opinion in genetics \& development 20(6):670--676

\bibitem[{Radulescu et~al(2008)Radulescu, Gorban, Zinovyev, and
  Lilienbaum}]{Radulescu-etal-2008}
Radulescu O, Gorban AN, Zinovyev A, Lilienbaum A (2008) Robust simplifications
  of multiscale biochemical networks. BMC Systems Biology 2:86,
  \doi{10.1186/1752-0509-2-86}

\bibitem[{Rand(2008)}]{Rand-2008}
Rand DA (2008) Mapping global sensitivity of cellular network dynamics:
  sensitivity heat maps and a global summation law. J R Soc Interface 5 Suppl
  1:S59--S69

\bibitem[{Saez-Rodriguez et~al(2004)Saez-Rodriguez, Kremling, Conzelmann,
  Bettenbrock, and Gilles}]{Saez-Rodriguez-2004}
Saez-Rodriguez J, Kremling A, Conzelmann H, Bettenbrock K, Gilles ED (2004)
  Modular analysis of signal transduction networks. Ieee Control Systems
  Magazine 24(4):35--52

\bibitem[{Schneider and Wilhelm(2000)}]{Schneider-2000}
Schneider KR, Wilhelm T (2000) Model reduction by extended quasi-steady-state
  approximation. Journal of Mathematical Biology 40(5):443--450

\bibitem[{Segel and Slemrod(1989)}]{Segel-Slemrod-1989}
Segel LA, Slemrod M (1989) The quasi-steady-state assumption: A case study in
  perturbation. SIAM Review 31:446--477

\bibitem[{Suckley and Biktashev(2003)}]{Suckley-Biktashev-2003}
Suckley R, Biktashev V (2003) Comparison of asymptotics of heart and nerve
  excitability. Physical Review E 68:011,902

\bibitem[{Sung et~al(2009)Sung, Salvatore, De~Lorenzi, Indrawan, Pasparakis,
  Hager, Bianchi, and Agresti}]{Sung-2009}
Sung MH, Salvatore L, De~Lorenzi R, Indrawan A, Pasparakis M, Hager GL, Bianchi
  ME, Agresti A (2009) Sustained oscillations of nf-kappab produce distinct
  genome scanning and gene expression profiles. PLoS One 4(9):e7163

\bibitem[{Tay et~al(2010)Tay, Hughey, Lee, Lipniacki, Quake, and
  Covert}]{Tay-etal-2010}
Tay S, Hughey JJ, Lee TK, Lipniacki T, Quake SR, Covert MW (2010) Single-cell
  nf-kappab dynamics reveal digital activation and analogue information
  processing. Nature 466(7303):267--271

\bibitem[{Tikhonov(1952)}]{Tikhonov-1952}
Tikhonov AN (1952) Systems of differential equations with small parameters at
  the derivatives. USSR Mathematics Sbornik 31(3):575--586

\bibitem[{Tur{\'a}nyi et~al(1993)Tur{\'a}nyi, Tomlin, and
  Pilling}]{Turanyi-etal-1993}
Tur{\'a}nyi T, Tomlin AS, Pilling MJ (1993) On the error of the
  quasi-steady-state approximation. J Phys Chem 97:163--172

\bibitem[{Turner et~al(2010)Turner, Paszek, Woodcock, Nelson, Horton, Wang,
  Spiller, Rand, White, and Harper}]{Turner-2010}
Turner DA, Paszek P, Woodcock DJ, Nelson DE, Horton CA, Wang Y, Spiller DG,
  Rand D, White MR, Harper CV (2010) Physiological levels of tnfalpha
  stimulation induces stochastic dynamics of nf-kappab respose in single living
  cells. J Cell Sci

\bibitem[{{Vasil'eva}(1952)}]{Vasilieva-1952}
{Vasil'eva} AB (1952) On differential equations containing small parameters.
  Mat Sb (NS) 31(73)(3):587--644

\bibitem[{Volpert and Hudjaev(1985)}]{Volpert-Hudjaev-1985}
Volpert AI, Hudjaev SI (1985) Analysis in classes of discontinuous functions
  and the equations of mathematical physics. Nijhoff, Dordrecht

\bibitem[{Wang et~al(2012)Wang, Paszek, Horton, Yue, White, Kell, Muldoon, and
  Broomhead}]{Wang-etal-2012}
Wang Y, Paszek P, Horton CA, Yue H, White MR, Kell DB, Muldoon MR, Broomhead DS
  (2012) A systematic survey of the response of a model nf-$\kappa$b signalling
  pathway to tnf$\alpha$ stimulation. Journal of theoretical biology
  297:137--147

\bibitem[{Whiteley(2010)}]{Whiteley-2010}
Whiteley JP (2010) Model reduction using a posteriori analysis. Mathematical
  Biosciences 225(1):44--52

\bibitem[{Yablonskii et~al(1991)Yablonskii, Bykov, Gorban, and
  Elokhin}]{Yablonskii-etal-1991}
Yablonskii GS, Bykov VI, Gorban AN, Elokhin VI (1991) Kinetic Models of
  Catalytic Reactions, Comprehensive Chemical Kinetics, vol~32. Elsevier,
  Amsterdam

\bibitem[{Zagaris et~al(2004)Zagaris, Kaper, and Kaper}]{Zagaris-etal-2004}
Zagaris A, Kaper HG, Kaper TJ (2004) Fast and slow dynamics for the
  computational singular perturbation method. Multiscale Model Simul
  2(4):613--638

\end{thebibliography}

\appendix

\section{Simplified Model vs Ashall model}
\label{app:SM}
\begin{figure}[tbp]
\centerline{\includegraphics[width=1.2\textwidth]{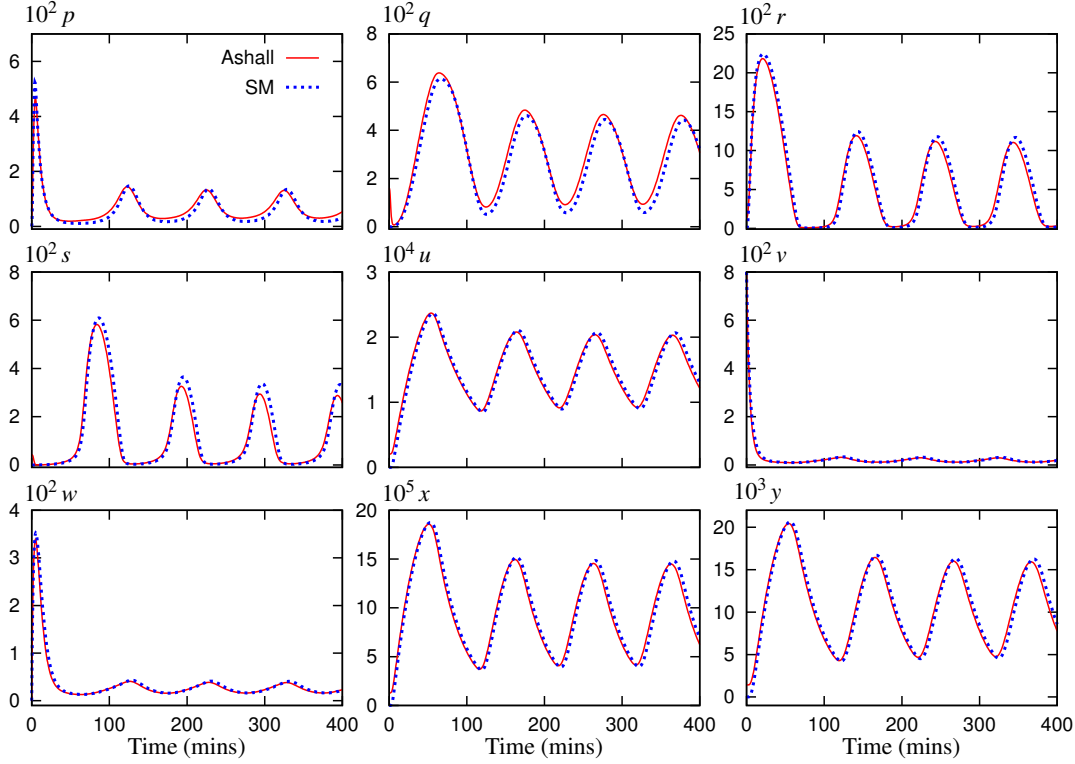}}
\caption{
Ashall vs Simplified Model. Time courses of solutions to the full
Ashall model and the Simplified Model, in response to continuous
TNF$\alpha$ treatment for time $t\geq0$ ($t$ in minutes). Clearly the
Simplifed Model gives close agreement with the full model, in terms of
variable amplitudes and the period of the limit cycle.
}
\label{fig:ashall_vs_SM}
\end{figure}

In figure~\ref{fig:ashall_vs_SM}, we show time courses of solutions to the full Ashall model and the Simplified Model, in response to continuous TNF$\alpha$ treatment, demonstrating the close agreement between the two models.

\section{Equations for the $z_0p_1y_1v_1s_1w_1$ model}

The model consists of four ordinary differential equations
\begin{subequations}                              \label{fmm}
  \begin{align}
\Df{\IkBa}{t} & = 
  -\kaoa\IkBa\bar{p}(\IkBa,\nNFkB,\tA)
  +\cta\tIkBa
  -\cfa\IkBa
  -\kitha\IkBa
  +\ketha\bar{s}(\IkBa,\nNFkB)
  -\kcoa\bar{w}(\nNFkB,\tA)\IkBa ,
  \label{fmm1} \displaybreak[2]\\
\Df{\nNFkB}{t} &= 
  \kio\kv\bar{p}(\IkBa,\nNFkB,\tA)
  -\kaoa\bar{s}(\IkBa,\nNFkB)\nNFkB
  -\keo\kv\nNFkB ,
  \label{fmm2} \displaybreak[2]\\
\Df{\tIkBa}{t} & =
  \coa\frac{\nNFkB^{h}}{\nNFkB^{h}+k^{h}}
  -\ctha\tIkBa ,
  \label{fmm3} \displaybreak[2]\\
\Df{\tA}{t} & =
  \co\frac{\nNFkB^{h}}{\nNFkB^{h}+k^{h}}
  -\cth\tA ,
  \label{fmm4}
  \end{align}
  \end{subequations}
where the functions in the right-hand side are defined by
  \begin{subequations}
  \begin{align}
   w^{0}(\tA) &=\frac{
    \TR\ka\kp\kbA\cf\NF
  }{
    \left(\kp\kbA\cf+\TR\ka\kbA\cf+\TR^{2}\ka\ct\tA\right)\ki
  } , \displaybreak[2]\\
  w^{1}(\nNFkB,\tA)&= \frac{
    \TR^{3}\ka^{2}\kp\kbA\cf\NF\ct\left(\frac{\co\nNFkB^{h}}{\nNFkB^{h}+k^{h}}-\cth\tA\right)
  }{
    \ki^{2}\left(\kp\kbA\cf+\TR\ka\kbA\cf+\TR^{2}\ka\ct\tA\right)^{2}
  } , \displaybreak[3]\\
  \bar{w}(\nNFkB,\tA) &= w^{0}(\tA)+w^{1}(\nNFkB,\tA) ,
  \label{fmm5} \displaybreak[2]\\
  \bar{y}(\nNFkB,\tA)
  &=\frac{\ct\tA}{\cf}-\frac{\cf(\co\frac{\nNFkB^{h}}{\nNFkB^{h}+k^{h}}-\cth\tA)}{\ct^{2}} ,
  \label{fmm6}\displaybreak[3]\\
 v^{0}(\nNFkB,\tA) &= \frac{
 \NF\kp\kbA
 }{
 \kp\kbA+\kbA\ka\TR+\TR^{2}\ka\bar{y}(\nNFkB,\tA)
 } , \displaybreak[2]\\ 
 v^{1}(\nNFkB,\tA) &= \frac{
 -\kp^{2}\TR\ct\NF\kbA^{2}(\co\frac{\nNFkB^{h}}{\nNFkB^{h}+k^{h}}-\cth\tA)
 }{
\cf(\frac{\kp\kbA}{\kbA+\TR\bar{y}(\nNFkB,\tA)}+\TR\ka)(\frac{\kp\kbA}{\kbA+\TR\bar{y}(\nNFkB,\tA)}-\TR\ka)^{2}(\kbA+\TR\bar{y}(\nNFkB,\tA))^{3}
 } , \displaybreak[2]\\
 \bar{v}(\nNFkB,\tA) &=v^{0}(\nNFkB,\tA)+v^{1}(\nNFkB,\tA) ,
 \label{fmm7}\displaybreak[3]\\
s_{\alpha}(\IkBa) &= \kv\kitha\IkBa ,
\displaybreak[2]\\
s_{\beta}(\nNFkB) &= \kaoa\nNFkB+\kv\ketha+\cf ,
\displaybreak[2]\\
p^{0}(\IkBa,\nNFkB,\tA) &=\frac{
    \keo\nNFkB+\kcta\bar{w}(\nNFkB,\tA)\left(\NF-\frac{\nNFkB}{\kv}\right)
  }{
    \kaoa\IkBa+\kio+\kcta\bar{w}(\nNFkB,\tA) ,
  }\displaybreak[2]\\
s_{\delta}(\IkBa,\nNFkB,\tA) &=\kv\kio
p^{0}(\IkBa,\nNFkB,\tA)-\kv\keo\nNFkB ,
\displaybreak[2]\\
s_{\gamma}(\IkBa,\nNFkB,\tIkBa,\tA) &= -\kaoa\IkBa
p^{0}(\IkBa,\nNFkB,\tA)-\kitha\IkBa-\cfa\IkBa+\cta\tIkBa-\kcoa\bar{w}(\nNFkB,\tA)\IkBa ,
\displaybreak[2]\\
\bar{s}(\IkBa,\nNFkB,\tIkBa,\tA) &= \frac{
s_{\alpha}(\IkBa)s_{\beta}(\nNFkB)^{2}+\kaoa s_{\delta}(\IkBa,\nNFkB,\tA)s_{\alpha}(\IkBa)-\kv\kitha s_{\gamma}(\IkBa,\nNFkB,\tIkBa,\tA)s_{\beta}(\nNFkB)
}{
\kaoa^{2}s_{\alpha}(\IkBa)\nNFkB+s_{\beta}^{3}(\nNFkB)+\kv\kitha\ketha
s_{\beta}(\nNFkB)} ,
 \label{fmm8}\displaybreak[3]\\
\alpha_{p}(\nNFkB,\tA) &=
\kv\keo\nNFkB+\kcta\bar{w}(\nNFkB,\tA)\left(\kv\NF-\nNFkB\right) ,
\displaybreak[2]\\
\beta_{p}(\IkBa,\nNFkB,\tA) &= \kv\left(\kaoa\IkBa + \kio +
  \kcta\bar{w}(\nNFkB,\tA)\right) ,
\displaybreak[2]\\
q_{\alpha}(\IkBa,\nNFkB,\tIkBa,\tA) &= \cta\tIkBa - \kitha\IkBa +
\ketha\nIkBa - \cfa\IkBa - \kcoa\bar{w}(\nNFkB,\tA)\IkBa ,
\displaybreak[2]\\
q_{\beta}(\IkBa) &= -\kaoa\IkBa ,
\displaybreak[2]\\
r_{\alpha}(\IkBa,\nNFkB,\tIkBa,\tA) &=
-\nNFkB\left(\kaoa\bar{s}(\IkBa,\nNFkB,\tIkBa,\tA) + \keo\kv\right) ,
\displaybreak[2]\\
r_{\beta} &= \kv\kio ,
\displaybreak[2]\\
p_{\delta}(\nNFkB,\tA) &= \kv\left(\kaoa q_{\alpha}(\IkBa,\nNFkB,\tIkBa,\tA) +
  \kcta\left(\TR\ka\bar{v}(\nNFkB,\tA) -
    \ki\bar{w}(\nNFkB,\tA)\right)\right) ,
\displaybreak[2]\\
p_{\gamma}(\IkBa) &= \kv\kaoa q_{\beta}(\IkBa,\nNFkB,\tIkBa,\tA) ,
\displaybreak[2]\\
p_{\epsilon}(\IkBa,\nNFkB,\tIkBa,\tA) &= \keo\kv
r_{\alpha}(\IkBa,\nNFkB,\tIkBa,\tA)-\kcta\bar{w}(\nNFkB,\tA)
r_{\alpha}(\IkBa,\nNFkB,\tIkBa,\tA) , \\
&+
\kcta\left(\TR\ka\bar{v}(\nNFkB,\tA)-\ki\bar{w}(\nNFkB,\tA)\right)\left(\kv\NF
  - \nNFkB\right) ,
\displaybreak[2]\\
p_{\zeta}(\nNFkB,\tA) &=
\left(\kv\keo-\kcta\bar{w}(\nNFkB,\tA)\right)r_{\beta} , 
\displaybreak[2]\\
\bar{p}(\IkBa,\nNFkB,\tIkBa,\tA) &=
\frac{\beta_{p}^{2}\alpha_{p}+\alpha_{p}p_{\delta}-\beta_{p}p_{\epsilon}}{\beta_{p}^{3}-\alpha_{p}p_{\gamma}+\beta_{p}p_{\zeta}} .
\label{fmm9}
\end{align}
\end{subequations}

\section{Equations for the $z_0p_0y_0v_0$ model}

Dynamic equations for the 6-variable model, $z_0p_0y_0v_0$, reduced using a representative solution for continuous \TNFA\ stimulation ($\TR\equiv1$):

\begin{subequations}                              \label{apmm}
  \begin{align}
    \Df{\IkBa}{t} &= -\kaoa\IkBa\bar{p}+\cta\tIkBa-\cfa\IkBa-\kitha\IkBa+\ketha\nIkBa-\kcoa\bar{w}\IkBa
    \displaybreak[2]\\
    \Df{\nNFkB}{t} &= \kio\kv\bar{p}-\kaoa\nIkBa\nNFkB-\keo\kv\nNFkB
    \displaybreak[2]\\
    \Df{\nIkBa}{t} &= \kitha\kv\bar{p}-\kaoa\nIkBa\nNFkB-\cfa\nIkBa-\ketha\kv\nIkBa
    \displaybreak[2]\\
    \Df{\tIkBa}{t} &= \coa\frac{\nNFkB^{h}}{\nNFkB^{h}+k^{h}}-\ctha\tIkBa
    \displaybreak[2]\\
     \Df{\IKKa}{t} &= \TR\ka\IKKn-\ki\IKKa
    \displaybreak[2]\\
    \Df{\tA}{t} &= \co\frac{\nNFkB^{h}}{\nNFkB^{h}+k^{h}}-\cth\tA
    \displaybreak[2]\\
    \bar{p} &= \frac{\keo\nNFkB\kv+\kcta\bar{w}\NF\kv-\kcta\bar{w}\nNFkB}{\kv\left(\kaoa\IkBa+\kio+\kcta\bar{w}\right)}
    \displaybreak[2]\\
    \bar{y} &= \frac{\ct\tA}{\cf}
     \displaybreak[2]\\
   \bar{v} &= \frac{\IK\kp\kbA}{\kbA\ka\TR+\TR^2\ka\A+\kp\kbA}
  \end{align}
\end{subequations}

\section{Equations for K6, K3 and K4 models}
\label{app:krishna}

\subsection{Krishna full model (K6)}
The full Krishna~\citep{Krishna} 7-variable model for $N_{n}$ and $N$ (free nuclear and cytoplasmic \NFKB\, $I_{m}$ (\IKBA\ mRNA), $I_{n}$ and $I$ (free nuclear and cytoplasmic I$\kappa$B), $(NI)_{n}$ and $(NI)$ (nuclear and cytoplasmic \NFKB:\IKBA) is given as:
\begin{subequations}
\label{subeq:krishna_7_PQRetc}
\begin{align}
\Df{N_{n}}{t} & = k_{Nin}N - k_{fn}N_{n}I_{n} + k_{bn}(NI)_{n},   \\
\Df{I_{m}}{t} & = k_{t}N_{n}^{2} - \gamma_{m}I_{m},   \\
\Df{I}{t} & =  k_{tl}I_{m} - k_{f}N I + k_{b}(NI) - k_{Iin} I + k_{Iout}I_{n},  \\
\Df{N}{t} & = -k_{f}N I + (k_{b}+\alpha)(NI) - k_{Nin}N,   \\
\Df{(NI)}{t} & = k_{f}N I - (k_{b}+\alpha)(NI) + k_{NIout}(NI)_{n},   \\
\Df{I_{n}}{t} & = k_{Iin}I - k_{Iout}I_{n} - k_{fn}N_{n}I_{n} + k_{bn}(NI)_{n},   \\
\Df{(NI)_{n}}{t} & = k_{fn}N_{n}I_{n} - (k_{bn}+k_{NIout})(NI)_{n}.  
\end{align}
\end{subequations}
Note that this can be replaced by a 6-variable system by using conservation of \NFKB\ to eliminate $N$. Base parameter values used in~\citep{Krishna} are given in Table~\ref{tab:krishna_params}.
\begin{table}[h]
\centerline{\begin{tabular}{|l|l|l||l|l|l|}
    \hline 
    Parameter & Value & Units & Parameter & Value & Units \\ \hline\hline
    $k_{Nin}$ & 5.4   & min$^{-1}$   & $k_{fn}$  & 30	&	$\mu\text{M}^{-1}\text{min}^{-1}$	\\ \hline
    $k_{bn}$ &  0.03 & min$^{-1}$   & $k_{t}$  & 1.03	&	$\mu\text{M}^{-1}\text{min}^{-1}$	\\ \hline
    $\gamma_{m}$ & 0.017  & min$^{-1}$   & $k_{tl}$  & 0.24 &	min$^{-1}$		\\ \hline
    $k_{f}$ & 30  & $\mu\text{M}^{-1}\text{min}^{-1}$   & $k_{b}$  & 0.03	&		\\ \hline			
    $k_{Iin}$ & 0.018   & min$^{-1}$  & $k_{Iout}$  & 0.012	&	min$^{-1}$	\\ \hline			
    $\alpha$ & 1.05 IKK  & min$^{-1}$   & $k_{NIout}$  & 0.83	&	min$^{-1}$	\\ \hline			
    IKK & 0.5  & $\mu\text{M}$   & $N_{tot}$  & 1	&	$\mu\text{M}$	\\ \hline			
  \end{tabular}}
\caption{Parameter values for Krishna model.}
\label{tab:krishna_params}
\end{table}

\subsection{Krishna 3-variable model (K3)}
The Krishna~\citep{Krishna} reduced model has 3 variables, and is given (in their Supplementary Material) as follows:

\begin{subequations}
\label{subeq:K3}
{\small
\begin{align}
\Df{N_{n}}{t} & = k_{Nin}K_{I}\frac{N_{tot}-N_{n}}{K_{I}+I} - k_{Iin} \frac{I N_{n}}{\delta + N_{n}},   \\
\Df{I_{m}}{t} & =  k_{t}N_{n}^{2} - \gamma_{m} I_{m}, \\
\Df{I}{t} & =  k_{tl}I_{m} - \alpha \frac{N_{tot}-N_{n}}{K_{I}+I} I,
\end{align}
}
\end{subequations}
where
\begin{equation}
K_{I}=\frac{k_{b}+\alpha}{k_{f}}, \qquad \qquad K_{N}=\frac{k_{bn}+K_{NIout}}{k_{fn}}, \qquad \qquad \delta=\frac{K_{N}}{N_{tot}}.
\end{equation}

\subsection{New 4-variable model}
The 4-variable model, obtained by applying our speed coefficient algorithm to K6 (with $N$ eliminated), comes from first-order QSSA for $N_{n}$ followed by zeroth-order QSSA for $(NI)_{n}$. The resulting system for variables $I_{m},\;I,\;(NI)$ and $I_{n}$ is given by:
\begin{subequations}
\label{subeq:vadim_P1U0}
\begin{align}
\Df{I_{m}}{t} & = k_{t}N_{n}^{2} \qquad - \gamma_{m}I_{m},   \\
\Df{I}{t} & =  \left[k_{tl}I_{m} + k_{b}(NI) + k_{Iout}I_{n}\right] \qquad - \left\{ k_{f}\left[N_{tot}-(NI)-N_{n}-(NI)_{n}\right] + k_{Iin} \right\} I,  \\
\Df{(NI)}{t} & = \left[k_{f}I \left\{N_{tot}-N_{n}-(NI)_{n}\right\} + k_{NIout}(NI)_{n} \right] \qquad - (k_{f}I+k_{b}+\alpha) (NI),   \\
\Df{I_{n}}{t} & = \left[k_{Iin}I + k_{bn}(NI)_{n}\right] \qquad - (k_{Iout}+k_{fn}N_{n}) I_{n},
\end{align}
with
\begin{align}
N_{n} & = \frac{-b-\sqrt{b^{2}-4ac}}{2a},  \\
(NI)_{n} & = \frac{k_{fn}N_{n}I_{n}}{k_{bn}+k_{NIout}},
\end{align}
where
\begin{align}
N_{tot} & = N + N_{n} + (NI)_{n} + (NI), \\
a  & = k_{NIout}k_{fn}^{3}(k_{Nin}-k_{bn}) I_{n}^{2}, \\
b & = -\Big(k_{bn} + k_{NIout}\Big) \;\;\; \Big\{ I_{n}  k_{NIout}  k_{Nin}  k_{fn}^{2} + k_{Nin}^{3}  k_{NIout} + 2  k_{Nin}  k_{NIout}  k_{fn}^{2}  I_{n}^{2}  \nonumber \\
	 & + k_{NIout}  k_{Nin}^{2}  k_{f}  I + k_{NIout}  k_{Nin}  k_{fn}  I_{n}  k_{f}I + k_{NIout}  k_{fn}^{3}  I_{n}^{3}   \nonumber \\
	 & + 2  k_{Nin}^{2}  k_{NIout}  k_{fn}  I_{n} - k_{NIout} k_{Nin}  (NI)  I_{n}  k_{fn}^{2} + 2  k_{Nin}^{2}  k_{bn}  k_{fn}  I_{n} + k_{Nin}  k_{bn} 
      k_{fn}^{2}  I_{n}^{2}   \nonumber \\
	 & + k_{Nin}^{2}  k_{f}  I  k_{bn} - k_{Nin}  k_{fn}^{2}  I_{n}^{2} 
      k_{IouI_{n}} + k_{bn}  k_{fn}^{2}  I_{n}^{2}  k_{IouI_{n}} + k_{Nin}^{3}  k_{bn} + k_{Nin}  k_{fn}  
     I_{n}  k_{f}  I  k_{bn}   \nonumber \\
	 & + k_{Nin}^{3}  k_{fn}  I_{n} + k_{Nin}  k_{fn}^{3}  I_{n}^{3} + 2  
     k_{Nin}^{2}  k_{fn}^{2}  I_{n}^{2} + k_{Nin}  k_{fn}^{2}  I_{n}  k_{Iin}  I - k_{bn}  
     k_{fn}^{2}  I_{n}  k_{Iin}  I   \nonumber \\
	 & + k_{Nin}^{2}  k_{f}  I  k_{fn}  I_{n} + k_{Nin}  k_{fn} ^2  I_{n}^{2}  k_{f}  I \Big\}, \\
c & = -k_{Nin} \Big(k_{bn} + k_{NIout}\Big)^{2} \;\;\; \Big\{ -k_{Nin}^{2} + k_{Nin}^{2} (NI) - 2 k_{Nin} k_{fn} I_{n} + k_{Nin} (NI) k_{b}    \nonumber \\
	 & + k_{Nin} (NI) \alpha + 2 k_{Nin} (NI) k_{fn} I_{n} - k_{Nin} k_{f} I + k_{Nin} (NI) k_{f} 
     I    \nonumber \\
	 & + k_{fn} I_{n} kIo - k_{fn} (NI) I_{n} kIo + k_{fn} I_{n} (NI) k_{f} I + k_{fn} 
     (NI) k_{Iin} I - k_{f} I k_{fn} I_{n}    \nonumber \\
	 & + k_{fn} I_{n} (NI) k_{b} - k_{fn}^{2} I_{n} ^{2} + k_{fn} I_{n} (NI) \alpha + (NI) k_{fn}^{2} I_{n}^{2} - k_{fn} k_{Iin} I \Big\}.			
\end{align}

\end{subequations}

\end{document}